\documentclass[usenatbib]{aa}
\usepackage[varg]{txfonts}

\usepackage[T1]{fontenc}
\usepackage{ae,aecompl}

\def\figdir{{./mmramdAAfigs}}

\usepackage{graphicx}	% Including figure files
\usepackage{amsmath}	% Advanced maths commands
\usepackage{amssymb}	% Extra maths symbols
\usepackage{mathtools}
\usepackage{array}
\usepackage{color}
\usepackage{bm}%Bold math symbols
\usepackage{mathrsfs}
\usepackage{dcolumn}

\pdfpageattr{/Group <</S /Transparency /I true /CS /DeviceRGB>>}
\def\eps{\varepsilon}

\def\norm#1{\left\Vert#1\right\Vert}

\def\norm#1{\Vert#1\Vert}
\def\Frac#1#2{{{\displaystyle\strut#1}\over{\displaystyle\strut#2}}}
\def\frac#1#2{{{#1}\over{#2}}}

\def\m@th{\mathsurround=0pt}
\def\matrice#1{\left[\,\vcenter{\normalbaselines\m@th
		\ialign{\hfil$##$\hfil&&\quad\hfil$##$\hfil\crcr
			\mathstrut\crcr\noalign{\kern-\baselineskip}
			#1\crcr\mathstrut\crcr\noalign{\kern-\baselineskip}}}\,\right]}
\def\EQM#1{\vcenter{\normalbaselines\m@th
		\ialign{${\displaystyle ##}$\hfil&&\ ${\displaystyle ##}$\hfil\crcr
			\mathstrut\crcr\noalign{\kern-\baselineskip}
			\noalign{\smallskip}
			#1\crcr\mathstrut\crcr\noalign{\kern-\baselineskip}}}}

\def\tu{{\bf\tilde u}}

\def\bea{\begin{eqnarray}}
\def\eea{\end{eqnarray}}
\def\be{\begin{equation}}
\def\ee{\end{equation}}

\def\G{{\bf G}}

\def\tu{{\bf\tilde u}}

\def\u{{\bf u}}                                                             
\def\x{{\bf x}}

\def\cC{{\mathscr{C}}}

\def\e{{\rm e}}

\def\nnb{\nonumber\\}

\def\d{\mathrm{d}}

\def\o{\mathrm{o}}
\def\O{\mathrm{O}}

\def\i{\mathbf{i}}

\def\expo#1{\mathbf{e}^{#1}}

\usepackage{nicefrac}

\def\N{\mathbb{N}}
\def\Z{\mathbb{Z}}

\def\H{\mathcal{H}}
\def\K{\mathcal{K}}
\def\Loz{\Lambda_{1,0}}
\def\Ltz{\Lambda_{2,0}}

\def\DG{\Delta G}
\def\x{\mathfrak{\scriptstyle X}}
\def\X{\mathbf{\mathfrak{X}}}

\def\da{\delta\alpha}
\renewcommand{\epsilon}{\varepsilon}
\def\apz#1{\alpha_{0,#1}}
\def\Da{\Delta\alpha}

\def\G{\mathcal{G}}

\def\hH{\hat{\H}}
\def\hK{\hat{\K}}
\def\hL{\hat{\Lambda}}
\def\hG{\hat{G}}
\def\hC{\hat{C}}
\def\hx{\hat{x}}
\def\hGa{\hat{\Gamma}}
\def \hU{\hat{U}}
\def \hT{\hat{T}}
\def\I{\mathcal{I}}
\def \Cm{{C_{\mathrm{min}}}}
\def \cm{c_\mathrm{min}}
\def \cE{\mathscr{E}}
\def \Cc{C_c^{(0)}}
\def \Cce{C_c^{(1)}}

\def \acir{\alpha_{\mathrm{cir}}}
\def \lc#1#2#3{b^{(#2)}_{#1}(#3)}

\def \bM{\beta^{(\mathrm{MMR})}}

\title{AMD-stability in the presence of first-order mean motion resonances}

\author{
A. C. Petit, J. Laskar \and G. Boué
}

\titlerunning{AMD-stability}
\authorrunning{Petit, Laskar \& Bou\'e}

\institute{
	ASD/IMCCE, CNRS-UMR8028, Observatoire de Paris,  PSL Research University, UPMC, 77 Avenue Denfert-Rochereau, 75014 Paris, France\\
	\email{antoine.petit@obspm.fr}
}

\date{Accepted XXX. Received YYY; in original form ZZZ}

\abstract{The AMD-stability  criterion allows to discriminate between a-priori stable planetary systems and systems 
	for which the stability is not granted and needs further investigations. AMD-stability is based on the 
	conservation of the Angular Momentum Deficit (AMD) in the averaged system at all orders of averaging.
	While the AMD criterion is rigorous, the conservation of the AMD is only granted in absence of mean-motion resonances (MMR).
	Here we extend the AMD-stability criterion to take into account mean-motion resonances, and more specifically
	the overlap of first-order MMR. If the MMR islands overlap, the system will	experience generalized chaos leading to instability. 
	The Hamiltonian of two massive planets on coplanar quasi-circular orbits can be reduced to an integrable one 
	degree of freedom problem for period ratios close to a first-order MMR. We use the reduced Hamiltonian to
	derive a new overlap criterion for first-order MMR.
	This stability criterion unifies the previous criteria proposed in the literature and admits the criteria 
	obtained for initially circular and eccentric orbits as limit cases.
	We then improve the definition of AMD-stability to take into account the short term chaos generated by MMR overlap.
	We analyze the outcome of this improved definition of AMD-stability  on selected multi-planet systems from the Extrasolar Planets Encyclop\ae dia. }

\begin{document}

\maketitle

\section{Introduction}

The AMD-stability  criterion \citep{Laskar2000,Laskar2017} allows to discriminate between a-priori stable planetary systems 
and systems needing an in-depth dynamical analysis to ensure their stability.
The AMD-stability is based on the conservation of the angular momentum deficit \citep[AMD,][]{Laskar1997} 
in the secular system at all orders of averaging \citep{Laskar2000,Laskar2017}.
Indeed, the conservation of the AMD fixes an upper bound to the eccentricities.
Since the semi-major axes are constant in the secular approximation,
a low enough AMD forbids collisions between planets.
The AMD-stability criterion has been used to classify planetary systems based on the stability of their secular dynamics \citep{Laskar2017}.

However, while the analytical criterion developed in \citep{Laskar2017} does not depend on series expansions for small
masses or spacing between the orbits, the secular hypothesis does not hold for systems experiencing mean motion resonances (MMR).
Although a system with planets in MMR can be dynamically stable, chaotic behavior may result from the overlap of 
adjacent MMR, leading to a possible increase of the AMD and eventually to close encounters, collisions or ejections.
For systems with small orbital separations, averaging over the mean anomalies is thus impossible due to
the contribution of the first-order MMR terms.
For example, two planets in circular orbits very close to each other are AMD-stable,
however the dynamics of this system cannot be approximated by the secular dynamics.
We thus need to modify the notion of AMD-stability in order to take into account those configurations.

In studies of planetary systems architecture, a minimal distance based on the Hill radius \citep{Marchal1982} is often used as a 
criterion of stability \citep{Gladman1993,Chambers1996,Smith2009,Pu2015}.
However, \cite{Deck2013} suggested that stability criteria based on the MMR overlap are more accurate in characterizing the instability of the three-body planetary problem.

Based on the considerations of \cite{Chirikov1979} for the overlap of resonant islands, 
\cite{Wisdom1980} proposed a criterion of stability  for the first-order MMR overlap in the context of the restricted circular three-body problem.
This stability criterion defines a minimal distance between the orbits such that the first-order MMR overlap with one another.
For orbits closer than this minimal distance, the MMR overlapping induces chaotic behavior eventually leading to the instability of the system.

Wisdom showed that the width of the chaotic region in the circular restricted problem is proportional to the ratio of the planet mass to the star mass to the power $2/7$.
\cite{Duncan1989} confirmed numerically that orbits closer than the Wisdom's MMR overlap condition were indeed unstable.
More recently, another stability criterion was  proposed by \cite{Mustill2012} to take into account the planet's eccentricity.
\cite{Deck2013} improved the two previous criteria by developing the resonant Hamiltonian for two massive, 
coplanar, low-eccentricity planets and \cite{Ramos2015} proposed a criterion of stability taking into account the second-order MMR in the restricted three-body problem.

While Deck's criteria are in good agreement with numerical simulations \citep{Deck2013} and can be applied to the three-body planetary 
problem, the case of circular orbits is still treated separately from the case of eccentric orbits.
Indeed, the minimal distance imposed by the eccentric MMR overlap stability criterion vanishes with eccentricities and therefore cannot 
be applied to systems with small eccentricities.
In this case, \cite{Mustill2012} and \cite{Deck2013} use the criterion developed for circular orbits.
A unified stability criterion for first-order MMR overlap had yet to be proposed.

In this paper, we propose in Section \ref{sec.res_ham} a new derivation of the MMR overlap criterion based on the development of the three-body Hamiltonian by \cite{Delisle2012}.
We show in Section \ref{sec_ovcrit} how to obtain  a unified criterion of stability working for both initially circular and eccentric orbits.
In Section \ref{sec.C_cecc}, we then use the defined stability criterion to limit the region where the dynamics can be considered to be 
secular and adapt the notion of AMD-stability thanks to the new limit of the secular dynamics.
Finally we study in Section \ref{sec_class} how the modification of the AMD-stability definition affects the classification proposed in \citep{Laskar2017}.

\section{The resonant Hamiltonian}
\label{sec.res_ham}

The problem of two planets close to a first-order MMR on nearly circular and coplanar orbits can be reduced
to a one-degree-of-freedom system through a sequence of canonical transformations \citep{Wisdom1986,Henrard1986,Delisle2012,Delisle2014}.
We follow here the reduction of the Hamiltonian used in \citep{Delisle2012,Delisle2014}.

\subsection{Averaged Hamiltonian in the vicinity of a resonance}

Let us consider two planets of masses $m_1$ and $m_2$ orbiting a star of mass $m_0$ in the plane.
We denote  the positions of the planets, $\u_i$, and the associated canonical momenta in the heliocentric frame, $\tu_i$.
The Hamiltonian of the system is \citep{Laskar1995}
\begin{align}
\hH=& \Frac{1}{2}\sum_{i=1}^{2}\left(\Frac{\norm{\tu_i}^2}{m_i}-\G \Frac{m_0m_i}{u_i}\right)+\nnb
&\frac{1}{2}\frac{\norm{\tu_1+\tu_2}^2}{m_0}-\G\frac{m_1m_2}{\Delta_{12}}
\end{align}
where $\Delta_{12}=\norm{\u_1-\u_2}$, and $\G$ is the constant of gravitation.
$\hH$ can be decomposed into a Keplerian part $\hK$ describing the motion of the planets if they had no masses and a perturbation part $\eps\hH_1$ due to the influence of massive planets,
\begin{align}
\hK&=\sum_{i=1}^{2}\frac{1}{2}\frac{\norm{\tu_i}^2}{m_i}-\frac{\G m_0m_i}{u_i}\\
\eps\hH_1&=\frac{1}{2}\frac{\norm{\tu_1+\tu_2}^2}{m_0}-\frac{\G m_1m_2}{\Delta_{12}}.
\end{align}
The small parameter $\epsilon$ is defined as the ratio of the planet masses over the star mass
\begin{equation}
\eps =\frac{m_1+m_2}{m_0}.
\end{equation}
Let us denote the angular momentum,
\begin{equation}
\vec{\hG}=\sum_{i=1}^{2}\u_i\wedge\tu_i
\end{equation}
which is simply the sum of the two planets Keplerian angular momentum.
$\vec{\hG}$ is a first integral of the system.
 
Following \citep{Laskar1991}, we  express the Hamiltonian in terms of the Poincar\' e coordinates
\begin{align}
\hH&=\hK+\eps\hH_1(\hL_i,\hx_i,\bar{\hx}_i)\nnb
	&= - \sum_{i=1}^2 \frac{\mu^2 m_i^3}{2\hL_i^2} 
+  \epsilon\underset{k\in \Z^2}{\sum_{l,\bar l \in \N^2}}C_{l,\bar l,k}(\hL) \prod_{i=1}^2 \hx_i^{l_i}
   \bar{\hx}_i^{\bar{l}_i}\expo{\i  k_i\lambda_i},
\end{align}
where $\mu=\G m_0$ and for $i=1,2$,
\begin{align*}
&\hL_i=m_i\sqrt{\mu a_i}\\
&\hG_i=\hL_i\sqrt{1-e_i^2}\\
&\hC_i=\hL_i-\hG_i\\
&\hx_i=\sqrt{\hC_i}\expo{-\i\varpi_i}\\
&\lambda_i=M_i+\varpi_i.
\end{align*}
Here, $M_i$ corresponds to the mean anomaly, $\varpi_i$ to the longitude of the pericenter,
$a_i$ to the semi-major axis and $e_i$ to the eccentricity of the Keplerian orbit of the planet $i$.
$\hG_i$ is the Keplerian angular momentum of planet $i$.
We use the set of symplectic coordinates of the problem $(\hL_i,\lambda_i,\hC_i,-\varpi)$, or the canonically associated variables $(\hL_i,\lambda_i,\hx_i,-\i \bar \hx_i)$.
The coefficients $C_{l,\bar l,k}$ depend on $\hL$ and the masses of the bodies.
They are linear combinations of Laplace coefficients \citep{Laskar1995}.
As a consequence of angular momentum conservation, the d'Alembert rule gives a relation on the indices  of the  non-zero $C_{l,\bar l,k}$ coefficients
\be
\sum_{i=1}^2 k_i-l_i+\bar l_i =0.
\label{dalemb}
\ee

We study here a system with periods close to the first-order MMR $p:p+1$ with $p\in \N^*$.
For periods close to this configuration, we have ${-pn_1+(p+1)n_2\simeq0}$,
where $n_i=\mu^2m_i^3/\hL_i^3$ is the Keplerian mean motion of the planet $i$.

\subsubsection{Averaging over non-resonant mean-motions}
Due to the $p:p+1$ resonance, we cannot average on both mean anomalies independently.
Therefore, there is no conservation of $\hL_i$ as in the secular problem.
However, the partial averaging over one of the mean anomaly gives another first integral.
Following \citep{Delisle2012}, we consider the equivalent set of coordinates $(\hL_i,M_i,\hG_i,\varpi_i)$, and make the following change of angles
\begin{equation}
\left(\begin{array}{c}
\sigma\\
M_2
\end{array}\right) = 
\left(\begin{array}{cc}
-p &p+1\\
0 & 1
\end{array}\right) \left(\begin{array}{c}
M_1\\
M_2
\end{array}\right).
\end{equation}
The actions associated to these angles are
\begin{equation}
\left(\begin{array}{c}
\hGa_1\\
\hGa
\end{array}\right) = 
\left(\begin{array}{cc}
	-\frac{1}{p} &0\\
	\frac{p+1}{p} & 1
	\end{array}\right)
\left(\begin{array}{c}
\hL_1\\
\hL_2
\end{array}\right)=\left(\begin{array}{c}
\frac{-1}{p}\hL_1\\
\frac{p+1}{p}\hL_1+\hL_2
\end{array}\right).
\end{equation}
We can now average the Hamiltonian over $M_2$ using a change of variables close to the identity given by the Lie series method.
Up to terms of orders $\eps^2$, we can kill all the terms with indices not of the form $C_{l,\bar l,-jp,j(p+1)}$.
In order to keep the notations light, we do not change the name of the variables after the averaging.
We  also designate the remaining coefficients $C_{l,\bar l,-jp,j(p+1)}$ by the lighter expression $C_{l,\bar l,j}$.
Since $M_2$ does not appear explicitly in the remaining terms, 
\be {\hGa=\frac{p+1}{p}\hL_1+\hL_2}\ee
is a first integral of the averaged Hamiltonian.
The parameter $p\hGa$ is often designed as the spacing parameter \citep{Michtchenko2008} and has been used extensively in the study of the first-order MMR dynamics.

Expressed with the variables $(\hL,\lambda,\hx,\bar \hx)$, the Hamiltonian can be written
\begin{align}
\hH_{\mathrm{av}}=& - \sum_{i=1}^2 \frac{\mu^2 m_i^3}{2\hL_i^2} +\nnb
  &\epsilon{\underset{j\in \Z}{\sum_{l,\bar l \in \N^2}}}C_{l,\bar l,j}(\hL) \hx_1^{l_1}
\bar{\hx}_1^{\bar{l}_1}\hx_2^{l_2}\bar{\hx}_2^{\bar{l}_2}\expo{\i j ((p+1)\lambda_2-p\lambda_1)},
\label{Ham_av}
\end{align}
where we dropped the terms of order $\eps^2$.

\subsubsection{Poincare-like complex coordinates}

\cite{Delisle2012} used a change of the angular coordinates in order to remove the exponential in the second term of
eq.~(\ref{Ham_av}) and use $\hG$ and $\hGa$ as actions.
The new set of angles $(\theta_\Gamma,\theta_G,\sigma_1,\sigma_2)$ is defined as

\begin{equation}
\left(\begin{array}{c}
\theta_\Gamma\\
\theta_G\\
\sigma_1\\
\sigma_2
\end{array}\right) = 
\left(\begin{array}{cccc}
p&-p&0&0\\ 
-p&p+1&0&0\\
-p&p+1&1&0\\
-p&p+1&0&1
\end {array}
\right)\cdot\left(\begin{array}{c}
\lambda_1\\
\lambda_2\\
-\varpi_1\\
-\varpi_2
\end{array}\right).
\end{equation}
The conjugated actions are 
\begin{equation}
\left(\begin{array}{c}
\hGa\\
\hG\\
\hC_1\\
\hC_2
\end{array}\right) = 
\left(\begin{array}{cccc}
\frac{p+1}{p} & 1 & 0 & 0\\
1 & 1 & -1 & -1\\
0 & 0 & 1 & 0\\
0&0&0&1
\end{array}\right)\cdot\left(\begin{array}{c}
\hL_1\\
\hL_2\\
\hC_1\\
\hC_2
\end{array}\right).
\end{equation}
We define $\hat{\x}_i=\sqrt{\hC_i}\expo{\i\sigma_i}$, the complex coordinates associated to $(\hC_i,\sigma_i)$.
Since we have $\hat \x_i=\hx_i\expo{\i\theta_G}$, the terms of the perturbation in (\ref{Ham_av}) can be written
\begin{align}
\prod_{i=1}^{2}\hat{x}_i^{l_i}\bar{\hat{x}}_i^{\bar{l}_i} \expo{\i j\theta_G}&=	\prod_{i=1}^{2}\hat{\x}_i^{l_i}\bar{\hat{\x}}_i^{\bar{l}_i}\expo{\i (-l_i+\bar l_i +j)\theta_G}\nnb
&= \prod_{i=1}^{2}\hat{\x}_i^{l_i}\bar{\hat{\x}}_i^{\bar{l}_i};
\end{align}
the last equality resulting from the d'Alembert rule~(\ref{dalemb}).
$\hGa$ and $\hG$ are conserved and the averaged Hamiltonian no longer depends on the angles $\theta_\Gamma$~and~$\theta_G$ 
\begin{equation}
\hH_{\mathrm{av}}= - \sum_{i=1}^2 \frac{\mu^2 m_i^3}{2\hL_i^2} 
+  \epsilon{\underset{j\in \Z}{\sum_{l,\bar l \in \N^2}}}C_{l,\bar l,j}(\hL) \prod_{i=1}^2 \hat{\x}_i^{l_i} \bar{\hat{\x}}_i^{\bar{l}_i}.
\end{equation}
$\hL_1$ and $\hL_2$ can be expressed as functions of the new variables and we have
\begin{align}
	\hL_1 &=-p(\hC+\hG-\hGa)\\
	\hL_2 &=(p+1)(\hC+\hG) -p\hGa,
\end{align}
where $\hC=\hC_1+\hC_2$ is the total AMD of the system.
Up to the value of the first integrals $\hGa$ and $\hG$, the system now has two effective degrees of freedom.

\subsection{Computation of the perturbation coefficients}

We now truncate the perturbation, keeping only the leading-order terms. Since we consider the first-order MMR,
the Hamiltonian contains some linear terms in $\x_i$.
Therefore the secular terms are neglected since they are at least quadratic.
Moreover, the restriction to the planar problem is justified since the inclination terms are at least of order two. 

We follow the method described in \cite{Laskar1991} and \cite{Laskar1995} to determine the expression of the perturbation $\hH_1$.
The details of the computation are given in Appendix \ref{res_ham_appendix}. Since we compute an expression at
first order in eccentricities and $\eps$, the semi major axis and in particular their ratio,
\be\alpha=\frac{a_1}{a_2},\ee are evaluated at the resonance.
At the first order, the perturbation term $\hH_1$ has for expression

\begin{equation}
\eps\hH_1=\hat R_1(\hat \x_1+\bar{\hat \x}_1)+\hat R_2(\hat\x_2+\bar{\hat\x}_2),
\end{equation}
where
\begin{align}
\hat R_1&= -\epsilon\frac{\gamma}{1+\gamma}\frac{\mu^2m_2^3}{\hL_2^2}\frac{1}{2}\sqrt{\frac{2}{\hL_1}}r_1(\alpha)\\
\mathrm{and}&\nnb
\hat R_2&= -\epsilon\frac{\gamma}{1+\gamma}\frac{\mu^2m_2^3}{\hL_2^2}\frac{1}{2}\sqrt{\frac{2}{\hL_2}}r_2(\alpha)\\
\end{align}
with $\gamma=m_1/m_2$, 
\begin{align}
r_1(\alpha)&=-\frac{\alpha}{4}\left(3\lc{3/2}{p}{\alpha}-2\alpha\lc{3/2}{p+1}{\alpha}-\lc{3/2}{p+2}{\alpha}\right),
 \label{coef_r1} \\
\mathrm{and\ \ }&\nnb
r_2(\alpha)&=\frac{\alpha}{4}\left(3\lc{3/2}{p-1}{\alpha}-2\alpha\lc{3/2}{p}{\alpha}-\lc{3/2}{p+1}{\alpha}\right)\nnb
           &\quad +\frac{1}{2}\lc{1/2}{p}{\alpha}.
           \label{coef_r2}
\end{align}
In the two previous expressions, $\lc{s}{k}{\alpha}$ are the Laplace coefficients that can be expressed as
\begin{equation}
\lc{s}{k}{\alpha}=\frac{1}{\pi}\int_{-\pi}^{\pi}\frac{\cos(k\phi)}{\left(1-2\alpha\cos\phi+\alpha^2\right)^s}\d\phi
\label{Lapcoef}
\end{equation}
for $k>0$. For $k=0$, a $1/2$ factor has to be added in the second-hand member of (\ref{Lapcoef}).

For $p=1$, it should be noted that a contribution from the kinetic part should be added \citep[Appendix \ref{res_ham_appendix} and][]{Delisle2012}
\be
\H_{1,i}=\frac{\mu^2 m_1^2m_2^2}{2m_0\hL_1\hL_2}\sqrt{\frac{2}{\hL_2}}(\hat \x_2+\bar{\hat \x}_2).
\ee

Using the expression of $\alpha$ at the resonance ${p:p+1}$,
\begin{equation}
\alpha_0=\left(\frac{p}{p+1}\right)^{2/3},
\end{equation}
we can give the asymptotic development of the coefficients $r_1$ and $r_2$ for $p\rightarrow +\infty$ (see Appendix~\ref{Res_coeff}).
The equivalent is
\begin{equation}
-r_1 \sim r_2 \sim \frac{K_1(2/3)+2K_0(2/3)}{\pi}(p+1).
\label{r_equi}
\end{equation}
where $K_\nu(x)$ is  the modified Bessel function of the second kind.
We note $r$ the numerical factor of the equivalent (\ref{r_equi}), we have
\begin{equation}
	r= \frac{K_1(2/3)+2K_0(2/3)}{\pi}= 0.80199 .
	\label{r_num}
\end{equation}
For the resonant coefficients $r_1$ and $r_2$, \cite{Deck2013} used  the  expressions $f_{p+1,27}(\alpha)$ and 
$f_{p+1,31}(\alpha)$ given in \cite[][pp. 539-556]{Murray1999}.
The expressions (\ref{coef_r1}) and (\ref{coef_r2}) are similar to $f_{p+1,27}(\alpha)$ and $f_{p+1,31}(\alpha)$ up 
to algebraic transformations using the relations between Laplace coefficients \citep{Laskar1995}.
In their computations, Deck et al. used a numerical fit of the coefficients for $p=2$ to 150 and obtained 
\be-f_{p+1,27}\sim f_{p+1,31}\sim0.802p.\ee
We obtain the same numerical factor $r$  through the analytical development of the functions $r_1$ and $r_2$. 

\subsection{Renormalization}

So far, the Hamiltonian has two degrees of freedom $(\hat \x_1,\bar{ \hat \x}_1,\hat \x_2,\bar{\hat \x}_2)$ and depends on two parameters $\hG$ and $\hGa$.
As shown in \citep{Delisle2012}, the constant $\hGa$ can be used to scale the actions, the Hamiltonian and the time without modifying the dynamics.
We define 
\begin{align*}
	\Lambda_i&=\hL_i/\hGa,\\
	G&=\hG/\hGa,\\
	C_i&=\hC_i/\hGa,\\
	\x_i&=\hat \x_i/\sqrt{\hGa},\\
	\H&=\hGa^2\hH,\\
	t&=\hat t/\hGa^3.
\end{align*}
With this change of variables, the new Hamiltonian no longer depends on $\hGa$.

The shape of the phase space is now only dependent on the first integral $G$.
However, $G$ does not vanish for the configuration around which the Hamiltonian is developed: the case of two resonant planets on circular orbits.
To be able to develop the Keplerian part in power of the system's parameter, we define $\DG=G_0-G$,  the difference in angular momentum between the circular resonant system and the actual configuration.
We have
\begin{equation}
G_0=\Loz+\Ltz,
\end{equation}
where $\Loz$ and $\Ltz$ are the value of $\Lambda_1$ and $\Lambda_2$ at resonance.
By definition, we have
\begin{equation}
\frac{\Loz}{\Ltz}=\gamma\left(\frac{p}{p+1}\right)^{1/3}=\gamma\sqrt{\alpha_0}.
\end{equation}
Moreover, we can express $\Loz$ as a function of the ratios $\alpha_0$ and $\gamma$,
\begin{equation}
\Loz=\frac{\hat\Lambda_{1,0}}{\hat\Gamma_0}	=\frac{1}{\left(\frac{p+1}{p}\right)+\frac{\Ltz}{\Loz}} = \left(\frac{p}{p+1}\right)\frac{\gamma}{\gamma+\alpha_0}.
\end{equation}
Similarly, $\Ltz$ can be expressed as
\begin{equation}
\Ltz=\frac{\alpha_0}{\alpha_0+\gamma}.
\end{equation}

Since $G_0$ is constant, $\DG$ is also a first integral of $\H$.
From now on, we consider $\DG$ as a parameter of the two-degrees-of-freedom $(\x_1,\x_2)$ Hamiltonian~$\H$.
The Keplerian part depends on the coordinates $\x_i$ through the dependence of $\Lambda_i$ in $C$.

$\Lambda_1$ and $\Lambda_2$ can be expressed as functions of the Hamiltonian coordinates and their value at the resonance,
\begin{align}
\Lambda_1&=\Loz-p(C-\DG)\nnb
\Lambda_2&=\Ltz+(p+1)(C-\DG).
\label{Li}
\end{align}

\subsection{Integrable Hamiltonian}

The system can be made integrable by a rotation of the coordinates $\x_i$ \citep{Sessin1984,Henrard1986,Delisle2014}.
We introduce $R$ and $\phi$ such that
\begin{equation}
R_1=R\cos(\phi) \quad \mathrm{and} \quad  R_2=R\sin(\phi).
\end{equation}
We have $R^2=R_1^2+R_2^2$ and $\tan(\phi)=R_2/R_1$. If we note $\mathcal{R}_\phi$ the rotation of angle $\phi$
we define $y$ such that $\x = \mathcal{R}_\phi y$.
We still have $C=\sum y_i\bar{y}_i$ so the only change in the Hamiltonian is the perturbation term
\begin{align}
\H&=\K(C,\DG)+R(y_1+\bar{y}_1)\nnb
&=\K(C,\DG)+2R\sqrt{I_1}\cos(\theta_1),
\end{align}
where $(I,\theta)$ are the  action-angle coordinates associated to $y$. With these coordinates, $I_2$ is a first integral. 
$R$ has for expression

\begin{equation}
R^2=\left(\frac{\eps\gamma}{1+\gamma}\frac{\mu^2m_2^3}{\Ltz^2}\right)^2
\left(\frac{r_1(\alpha_0)^2}{2\Loz}+\frac{r_2(\alpha_0)^2}{2\Ltz}\right).
\end{equation}

We now develop the Keplerian part around the circular resonant configuration in series of $(C-\DG)$
thanks to the relations~(\ref{Li}).
We develop the Keplerian part to the second order in $(C-\DG)$ since the first order vanishes (see Appendix~\ref{dev_kep}).
The computation of the second-order coefficient gives
\begin{equation}
\frac{1}{2}\K_2=-\frac{3}{2}\mu^2m_2^3\frac{(\gamma+\alpha_0)^5}{\gamma \alpha_0^4}(p+1)^2.
\end{equation}
We drop the constant part of the Hamiltonian and obtain the following expression
\begin{equation}
\H=\frac{\K_2}{2}(I_1 +I_2 -\DG)^2 +2R\sqrt{I_1}\cos(\theta_1).
\end{equation}
We again change the time scale by dividing the Hamiltonian by $-\K_2$ and multiplying the time by this factor.
We define 
\be\chi=-\frac{\sqrt{2}R}{\K_2}
\label{chi_def}\ee
and after simplification,
\begin{align}
\chi&=\frac{1}{3}\frac{\eps(\gamma\alpha_0)^{3/2}}{(1+\gamma)(\alpha_0+\gamma)^2}\frac{r_2(\alpha_0)}{(p+1)^2}f(p)\\
    &=\frac{r}{3}\frac{\eps\gamma^{3/2}}{(1+\gamma)^3}\frac{1}{p+1}+\O((p+1)^{-2}),
\end{align} 
where $r$ was defined in (\ref{r_num}) and ${f(p)=1+\O(p^{-1})}$ is a function of $p$ and $\gamma$
\begin{equation}
f(p)=\sqrt{1-\frac{\alpha_0}{\alpha_0+\gamma}\left(1-\frac{p+1}{p}\left(\frac{r_1}{r_2}\right)^2\right)}.
\end{equation}
At this point the Hamiltonian can be written
\begin{equation}
\H=-\frac{1}{2}(I_1+I_2-\DG)^2+\chi\sqrt{2I_1}\cos(\theta_1)
\end{equation}
and has almost its final form. 
We divide the actions and the time by $\chi^{2/3}$ and the Hamiltonian by $\chi^{4/3}$ and we obtain
\begin{equation}
\H_A=-\frac{1}{2}(\I-\I_0)^2-\sqrt{2\I}\cos(\theta_1),
\label{ham_I}
\end{equation}
where
\begin{equation}
\I=\chi^{-2/3}I_1\quad \mathrm{and}\quad \I_0=\chi^{-2/3}(\DG-I_2).
\label{I}
\end{equation} 

\subsection{Andoyer Hamiltonian}

\begin{figure}
	\begin{center}
		\includegraphics[width=8cm]{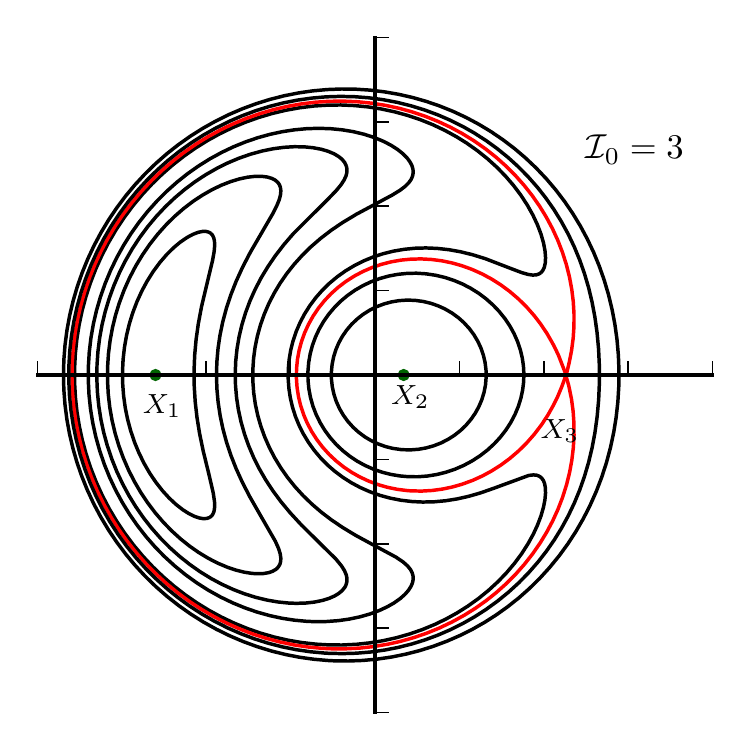}
		\caption{Hamiltonian $\H_A$ (\ref{ham_And}) represented with the saddle point and the separatrices in red.}
		\label{fig.hamiltonian}
	\end{center}
\end{figure}

We now perform a polar to Cartesian change of coordinates with 
\begin{align}
X&=-\sqrt{2\I}\cos(\theta_1),\nnb
Y &=\sqrt{2\I}\sin(\theta_1).
\label{Itet_to_XY}
\end{align}
We change the sign of $X$ in order to have the same orientation as \citep{Deck2013}.
Doing so, the Hamiltonian becomes
\begin{equation}
\H_A=-\frac{1}{2}\left(\frac{1}{2}(X^2+Y^2)-\I_0\right)^2 -X.
\label{ham_And}
\end{equation}
We recognize the second fundamental model of resonance \citep{Henrard1983}.
This Hamiltonian is also called an Andoyer Hamiltonian \citep{Ferraz-Mello2007}.
We show in Figure \ref{fig.hamiltonian} the level curves of the Hamiltonian $\H_A$ for $\I_0=3$.

The fixed points of the Hamiltonian satisfy the equations
\begin{align}
	\dot{X}&=Y\left(\frac{1}{2}(X^2+Y^2)-\I_0\right)=0\\
	\dot{Y}&=-X\left(\frac{1}{2}(X^2+Y^2)-\I_0\right) -1=0,
\end{align}
which have for solutions $Y=0$ and the real roots of the cubic equation in $X$
\begin{equation}
X^3-2\I_0X+2=0.
\label{cubiceq}
\end{equation}
Equation~(\ref{cubiceq}) has three solutions \citep{Deck2013} if its determinant $\Delta=32(\I_0^3-27/8)>0$, \emph{i.e.} $\I_0>3/2$.
In this case, we note these roots $X_1<X_2<X_3$.
$X_1$ and $X_2$ are elliptic fixed points while $X_3$ is a hyperbolic one.

\section{Overlap criterion}
\label{sec_ovcrit}
As seen in the previous section, the motion of two planets near a first-order MMR can be reduced to an integrable system for small eccentricities and planet masses.
However, if two independent combinations of frequencies are close to zero at the same time, the previous reduction is not valid anymore.
Indeed, we must then keep, in the averaging, the terms corresponding to both resonances.
While for a single resonant term the system is integrable, overlapping resonant islands will lead to chaotic motion \citep{Chirikov1979}.

\cite{Wisdom1980} first applied the resonance overlap criterion to the first-order MMR and found,
in the case of the restricted three-body problem with a circular planet, that the overlap occurs for
\begin{equation}
	1-\alpha<1.3\eps^{2/7}.
\end{equation} 
Through numerical simulations, \citep{Duncan1989} confirmed Wisdom's expression up to the numerical coefficient ($1-\alpha<1.5\  \eps^{2/7}$).
A similar criterion was then developed by \cite{Mustill2012} for an eccentric planet, they found that for an
eccentricity above $0.2\ \eps^{3/7}$, the overlap region satisfies the criterion $1-\alpha<1.8(\eps\ e)^{1/5}$.
\cite{Deck2013} adapted those two criteria to the case of two massive planets, finding little difference up to the numerical coefficients.
However, they  treat  two different situations; the case of orbits initially circular and the case 
of two eccentric orbits.
As in \citep{Mustill2012}, the eccentric criterion proposed can be used for eccentricities verifying $e_1+e_2 \gtrsim 1.33\ \eps^{3/7}$.
We show here that the two Deck's criteria can be obtained as the limit cases of a general expression.

\subsection{Width of the libration area}
\label{sec.widthsma}
Using the same approach as \citep{Wisdom1980,Deck2013}, we have to express the width of the resonant island as a
function of the orbital parameters and compare it with the distance between the two adjacent centers of MMR.

In the ($X,Y$) plane, the center of the resonance is located at the point of coordinates $(X_1,0)$.
The width of the libration area is defined as the distance between the two separatrices on the $Y=0$ axis.
It is indeed the direction where the resonant island is the widest.

We note $X_1^*,X_2^*$ the abscissas of the intersections between the separatrices and the $Y=0$ axis.
Relations between $X_1^*,X_2^*$, and $X_3$  can be derived (see Appendix \ref{coef_poly}) and we obtain the expressions of $X^*_1$ and $X^*_2$ as functions of $X_3$ \citep{Ferraz-Mello2007,Deck2013}.
We have
\begin{align}
X_1^*&=-X_3-\frac{2}{\sqrt{X_3}},\\
X_2^*&=-X_3+\frac{2}{\sqrt{X_3}}.
\end{align}
The width of the libration zone $\delta X$ depends solely on the value of $X_3$, 
\begin{equation}
\delta X=\frac{4}{\sqrt{X_3}}.
\end{equation}

In order to study the overlap of resonance islands, we need the width of the resonance in terms of $\alpha$.
Let us invert the previous change of variables in order to express the variation of $\alpha$ in terms of the variation of $X$. 
In this subsection, for any function $Q(X)$, we note
 \be\delta Q=|Q(X_1^*)-Q(X_2^*)|.\ee
The computation of $\delta\I$ (\ref{Itet_to_XY}) is straightforward from the computation of $\delta X$
\begin{align}
\delta \I&=\left|\frac{{X_1^*}^2}{2}-\frac{{X_2^*}^2}{2}\right|\nnb
&=\frac{1}{2}\left|{X_2^*}+{X_1^*}\right|\left|{X_2^*}-{X_1^*}\right|\nnb
&=X_3\delta X\nnb
\delta \I&=4\sqrt{X_3}.
\end{align}
We then directly deduce $\delta I_1=\chi^{2/3}\delta \I$ from (\ref{I}). Since $I_2$ and $\DG$ are first integrals, the variation of $\Lambda_i$ only depends on $\delta I_1$. And finally, since we have
\begin{equation}
\alpha=\left(\gamma^{-1}\frac{\Lambda_1}{\Lambda_2}\right)^2,
\end{equation}
$\alpha$ can be developed to the first order in $(C-\DG)$ thanks to (\ref{Li}). This development gives
\begin{equation}
\alpha=\alpha_0\left(1-\frac{2(\alpha_0+\gamma)^2}{\gamma\alpha_0}(p+1)(I_1-\chi^{2/3}\I_0)\right).
\end{equation}
The width of the resonance in terms of $\alpha$ is then directly related to $X_3$ through
\begin{equation}
\delta \alpha=\alpha_0\frac{8r^{2/3}}{3^{2/3}}\eps^{2/3}(p+1)^{1/3}\sqrt{X_3}+\o(\eps^{2/3}(p+1)^{1/3}).
\label{width}
\end{equation}
The computation of the width of resonance is thus reduced to the computation of the root $X_3$ as a function of the parameters.
It should also be remarked that at the first order, the width of resonance does not depend on the mass ratio $\gamma$.

\subsection{Minimal AMD of a resonance}

We are now interested in the overlap of adjacent resonant islands.
Planets trapped in the chaotic zone created by the overlap will experience variations of their actions eventually leading to collisions.

For a configuration close to a given resonance $p:p+1$, the AMD can evolve toward higher values
if the original value places the system in a configuration above the inner separatrix, 
eventually leading the planets to collision or chaotic motion in case of MMR overlap.
On the other hand, if the initial AMD of the planets forces them to remain in the inner circulation region of the overlapped MMR islands, the system will remain stable in regards to this criterion.
Since $C=I_1+I_2$, and $I_2$ is a first integral, we define the minimal AMD  of a resonance\footnote{We summarize the notations of the various AMD expressions used in this paper in Table~\ref{table_not}.} $\Cm(p)$ as the minimal value of $I_1$ to enter the resonant island given $\DG-I_2$.
Two cases must be discussed:
\begin{itemize}
	\item The point $I_1=0$ is already in the libration zone and then $\Cm=0$,
	\item The point $I_1=0$ is in the inner circulation zone and then we have
\end{itemize}
\begin{equation}
\Cm=I_1(X_2^*)=\frac{\chi^{2/3}}{2}\left(X_3-\frac{2}{\sqrt{X_3}}\right)^2.
\label{Cmin_def}
\end{equation}
In the second case, we have an implicit expression of $X_3$ depending on $\Cm$
\begin{equation}
\chi^{-1/3}\sqrt{2\Cm}=X_3-\frac{2}{\sqrt{X_3}},
\label{impliX3}
\end{equation}
where $\chi$ was defined in (\ref{chi_def}).
In other words, there is a one-to-one correspondence between $\Cm$ (\ref{Cmin_def}) and the Hamiltonian parameter $\I_0$
for $\Cm>0$. The shape of the resonance island is completely described by $\Cm$.

We can also use the definition of $\Cm$ to give an expression depending on the system parameters
\begin{align}
\Cm&=I_1=u_1\bar{u}_1\nnb
&=\left|\frac{R_1}{R}\sqrt{\frac{\Loz}{2}}\X_1+\frac{R_2}{R}\sqrt{\frac{\Ltz}{2}}\X_2\right|^2\nnb
&=\left(\frac{R_1}{R}\right)^2\frac{\Loz}{2}\left|\X_1-\left|\frac{R_2}{R_1}\right|\sqrt{\frac{\Ltz}{\Loz}}\X_2\right|^2\nnb
&\simeq\frac{\alpha_0\gamma}{2(\alpha_0+\gamma)^2}(c_1^2+c_2^2-2c_1c_2\cos\Delta\varpi),
\label{Cmin}
\end{align}
where $c_i=\sqrt{2}\sqrt{1-\sqrt{1-e_i^2}}=|\X_i|$.
We note 
\begin{equation}
{\cm=c_1^2+c_2^2-2c_1c_2\cos\Delta\varpi,}
\label{cm}
\end{equation} 
the reduced minimal AMD.
We can use the expression~(\ref{Cmin}) to compute the quantity $\chi^{-1/3}\sqrt{2\Cm}$ appearing in equation~(\ref{impliX3})
\begin{equation}
\chi^{-1/3}\sqrt{2\Cm}\simeq\frac{3^{1/3}}{r^{1/3}}\frac{(p+1)^{1/3}}{\eps^{1/3}}\sqrt{\cm} +\o(p^{1/3}).
\label{Cm(cm)}
\end{equation}

\begin{figure}
	\begin{center}
		\includegraphics[width=7cm]{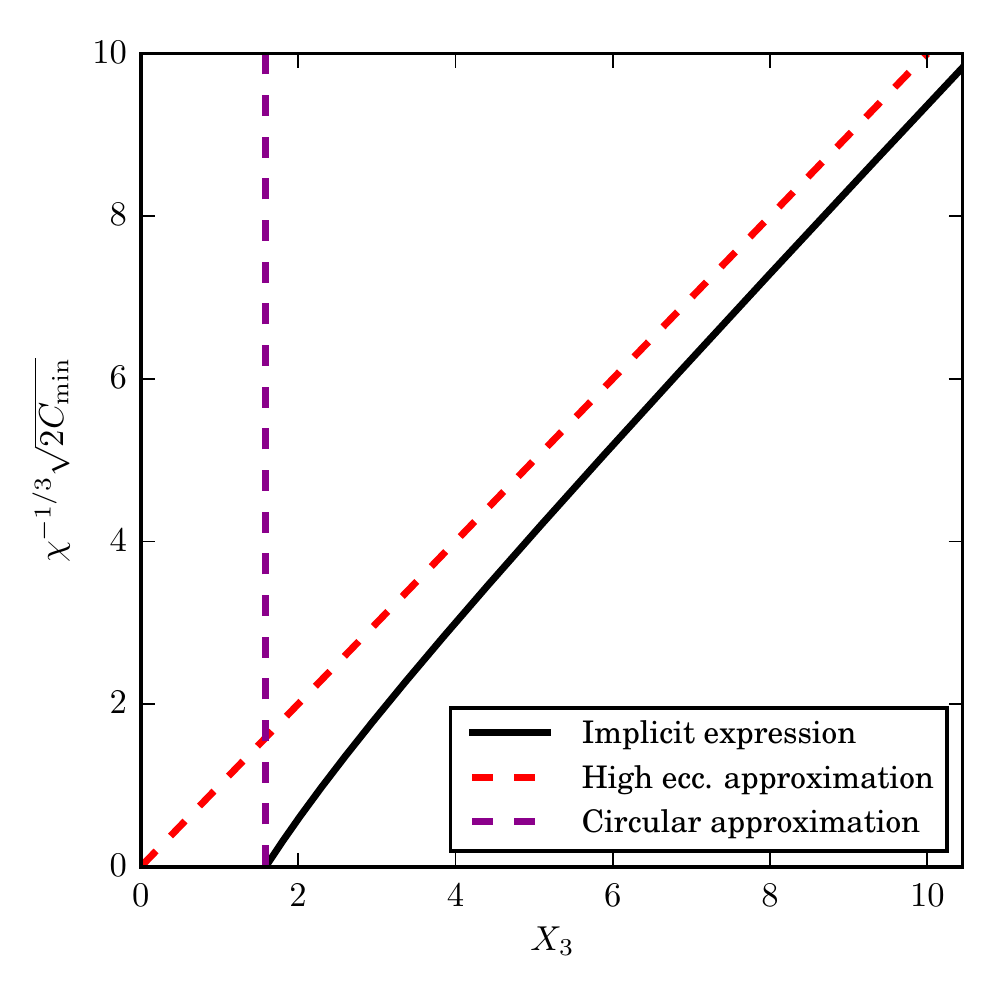}
		\caption{Relation (\ref{impliX3}) between $X_3$ and $\Cm$ (\ref{Cmin_def}) and two different approximations. In red, the approximation used by \cite{Deck2013} for eccentric orbits and in purple the constant evaluation used for circular orbits.}
		\label{figimpliX3}
	\end{center}
\end{figure}

The function $\Cm(X_3)$ (Eq. \ref{Cmin_def}) is plotted in Figure~\ref{figimpliX3} with the two approximations used by \cite{Deck2013} to obtain the width of the resonance. For $\Cm\gg\chi^{2/3}$ or $\Cm$ close to zero, the relation can be simplified and we obtain
\begin{align}
X_3&\sim\chi^{-1/3}\sqrt{2\Cm}\label{impX3_highecc}\\
X_3&=2^{2/3}+\frac{2}{3}\chi^{-1/3}\sqrt{2\Cm} +\O(\chi^{-2/3}\Cm)\label{impX3_lowecc}.
\end{align}
We can use the developments~(\ref{impX3_highecc}) and (\ref{impX3_lowecc}) in order to compute the width of the resonance in these two cases (see Appendix \ref{a.width}).
It should be noted as well that for $\Cm = 0$, we have $X_3=2^{2/3}$.

\begin{table*}
	\begin{center}
		\caption{Summary of the diverse notations of AMD used in this paper.}
		
		\begin{tabular}{lll}
			\hline
			Notation & Description & Equation\\
			\hline\vspace{-6pt}\\
			$C$ & Total AMD of the system & \\
			$\Cm$ & Minimal AMD to enter a resonance island & (\ref{Cmin_def}) \\
			$\cm$ & Normalized minimal AMD & (\ref{cm})\\
			$\cC$ & Relative AMD & (\ref{rel_AMD})\\
			$\Cc$ & Critical AMD deduced from the collision condition & \citep{Laskar2017}\\
			$\Cce$ & Critical AMD deduced from the MMR overlap & (\ref{Cce})\\
			$C_c$ & Complete critical AMD & (\ref{amd_crit})\\
			\hline
			
		\end{tabular}
		\label{table_not}
	\end{center}
	
\end{table*}

\subsection{Implicit overlap criterion}

The overlap of MMR can be determined by finding the first integer $p$ such that
the sum of the half-width of the resonances $p:p+1$ and $p+1:p+2$ is larger than the distance between
the respective centers of these two resonances \citep{Wisdom1980,Deck2013}
\begin{equation}
\frac{\Da}{\apz{p}}\lesssim \frac{1}{2}\left(\frac{\da_{p}}{\apz{p}}+\frac{\da_{p+1}}{\apz{p+1}}\right),
\label{crit_gen}
\end{equation}
where $\Da$ is the distance between the two centers and $\da_k$ corresponds to the width of the resonance $k:k+1$.

Up to terms of order $\eps^{2/3}$, the center of the resonance island $p:p+1$ is located at  the center of the resonance of the unperturbed Keplerian problem, ${\alpha_{0,p}=(p/(p+1))^{2/3}}$. 
We develop $\alpha_{0,p}$ for $p\gg1$
\begin{align}
\alpha_{0,p}&=\left(\frac{p}{p+1}\right)^{\nicefrac{2}{3}}\nnb
            &=1-\frac{2}{3(p+1)}-\frac{1}{9(p+1)^2}+\O((p+1)^{-3}).
\label{alpha_p}
\end{align} 
Therefore, we have at second order in $p$
\begin{equation}
\frac{\Da}{\alpha_{0,p}}=\frac{2}{3}\frac{1}{(p+1)^2}.
\label{Da}
\end{equation}

We can use the implicit expression~(\ref{impliX3}) of $X_3$ as a function of $\sqrt{\cm}$ (Eq. \ref{cm}) in order to 
derive an overlap criterion independent of approximations on the value of $\Cm$.
Equating the general width of resonance~(\ref{width}) with the distance between to adjacent centers~(\ref{Da}) and isolating $X_3$ gives
\begin{equation}
X_3=\frac{3^{4/3}}{144r^{4/3}}\epsilon^{-4/3}(p+1)^{-14/3}.
\end{equation}
We can inject this expression of $X_3$ into~(\ref{impliX3}), and using equation (\ref{Cm(cm)}),
\begin{equation}
\sqrt{\cm}=\frac{1}{48r\epsilon(p+1)^5}-8r\epsilon(p+1)^2.
\end{equation}
Using the first order expression of $(p+1)$ as a function of $\alpha$,
\begin{equation}
\frac{1}{p+1}=\frac{3}{2}(1-\alpha)
\end{equation}
we obtain an implicit expression of the overlap criterion
\begin{equation}
\sqrt{\cm}=\frac{3^4(1-\alpha)^5}{2^9r\epsilon}-\frac{32r\epsilon}{9(1-\alpha)^2}.
\label{impli_crit}
\end{equation}

\subsection{Overlap criterion for circular orbits}

The implicit expression~(\ref{impli_crit}) can be used to find the criteria proposed by \cite{Deck2013} for circular and eccentric orbits.
Let us first obtain the circular criterion by imposing $\cm=0$ in equation~(\ref{impli_crit})
\begin{equation}
	3^6(1-\alpha)^7=2^{14}r^2\epsilon^2.
\end{equation}
We can express $1-\alpha$ as a function of $\eps$ and we obtain
\begin{equation}
1-\alpha_{\mathrm{overlap}}=\frac{4 r^{2/7}}{3^{6/7}}\epsilon^{2/7}=1.46 \epsilon^{2/7}.
\label{circ_crit}
\end{equation}
The exponent 2/7 was first proposed by \cite{Wisdom1980} and the numerical factor $1.46$ is similar to the one found by \cite{Deck2013}.

\subsection{Overlap criterion for high-eccentricity orbits}

For large eccentricity, \cite{Deck2013} proposes a criterion based on the development~(\ref{impX3_highecc}) of equation~(\ref{impliX3}).
This criterion is obtained from~(\ref{impli_crit}) by ignoring the second term of the right-hand side which leads to
\begin{equation}
	2^9r\epsilon\sqrt{\cm}=3^4(1-\alpha)^5.
\end{equation}
Isolating $1-\alpha$ gives
\begin{equation}
1-\alpha=\frac{2^{9/5}}{3^{4/5}}r^{1/5} \epsilon^{1/5}\cm^{1/10}=1.38 \epsilon^{1/5}\cm^{1/10}.
\label{hecc_crit}
\end{equation}
This result is also similar to Deck's one.
For small $\cm$, the criterion~(\ref{hecc_crit}) is less restrictive than the criterion (\ref{circ_crit}) obtained for circular orbits.
The comparison of these two overlap criteria provides a minimal value of $\cm$ for the validity of the eccentric criterion
\begin{equation}
	\sqrt{\cm}=1.33\eps^{3/7}.
\end{equation}

\subsection{Overlap criterion for low-eccentricity orbits}

For smaller eccentricities, we can develop the equation~(\ref{impli_crit}) for small $\sqrt{\cm}$ and $\alpha$ close
to ${\acir=1-1.46\eps^{2/7}}$, the critical semi major axis ratio for the circular overlap criterion~(\ref{circ_crit}).
We have
\begin{equation}
3^22^9r\eps(1-\alpha)^2\sqrt{\cm}=3^6(1-\alpha)^7-2^{14}r^2\epsilon^2.
\end{equation}
We develop the right-hand side at the first order in $(\acir-\alpha)$ and evaluate the left-hand side for $\alpha=\acir$ and after some simplifications obtain
\begin{equation}
	\acir-\alpha=\frac{2^9r\eps}{7\times3^4}\frac{\sqrt{\cm}}{(1-\acir)^4}.
\end{equation}
We inject the expression of $\acir$ into this equation and obtain the following development of the overlap criterion for low eccentricity:
\begin{equation}
	\acir-\alpha=\frac{2\sqrt{\cm}}{7\times3^{4/7}r^{1/7}\epsilon^{1/7}}=0.157\frac{\sqrt{\cm}}{\epsilon^{1/7}}.
\end{equation}
This development remains valid for small enough $\sqrt{\cm}$ if ${\acir-\alpha\ll1-\acir}$, which can be rewritten
\begin{equation}
	0.157\eps^{-1/7}\sqrt{\cm}\ll1.46\eps^{2/7},
\end{equation}
which leads to
\begin{equation}
\sqrt{\cm}\ll 9.30 \eps^{3/7}.
\end{equation}
It is worth noting that the low-eccentricity approximation allows to cover the range of eccentricities where the criterion (\ref{hecc_crit}) is not applicable, since both boundaries depend on the same power of $\eps$.

\begin{figure*}
	\begin{center}
		\includegraphics[width=15cm]{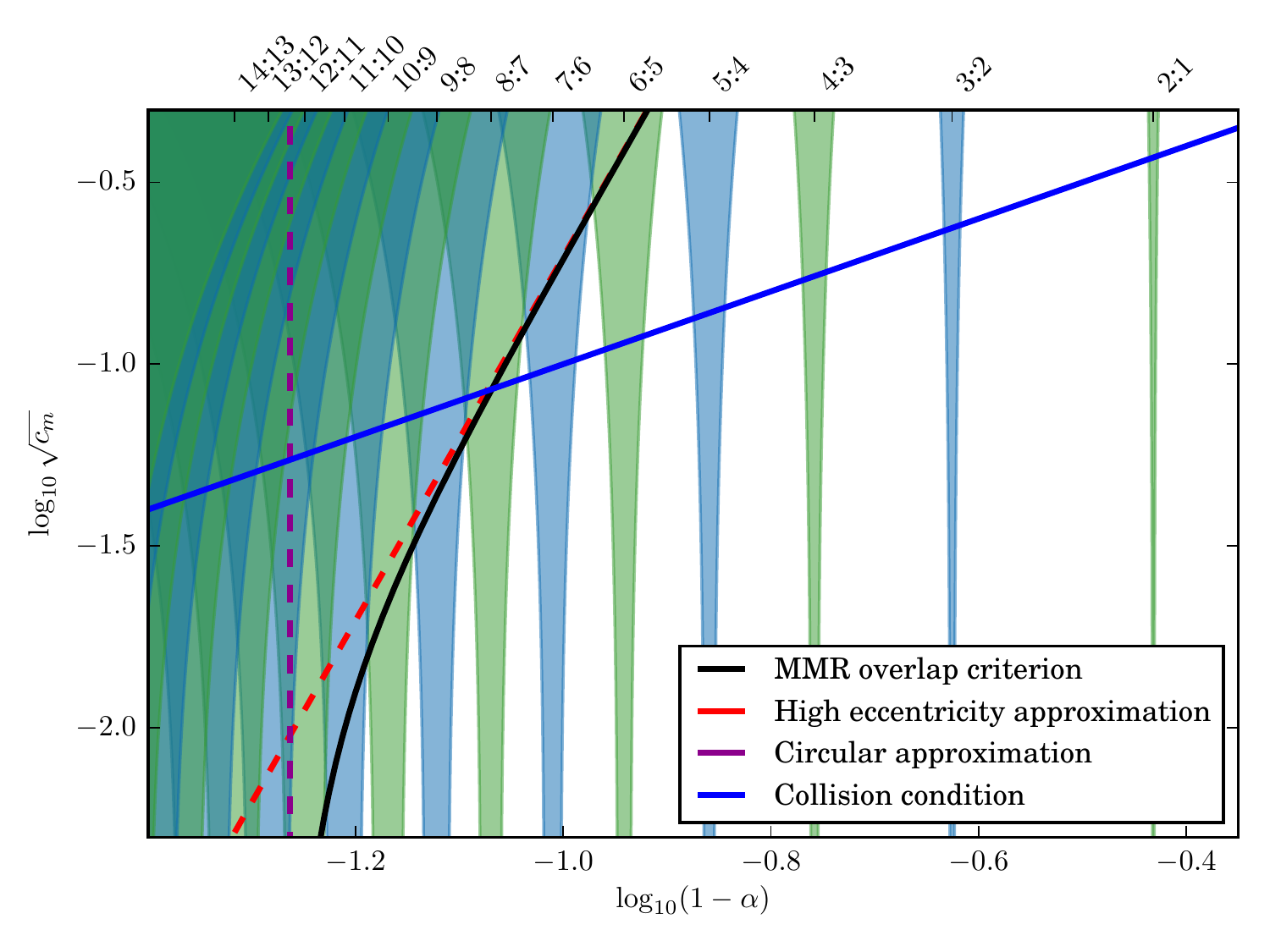}
		\caption{Representation of the MMR overlap criteria. The dotted lines correspond to the criteria proposed by
	\citep{Deck2013}, and the collision curve is the approximation of the collision curve for $\alpha\rightarrow1$.
We represented in transparent green ($p$ odd) and blue ($p$ even) the first $p:p+1$ MMR islands to show the agreement between the proposed overlap criterion and the actual intersections.
In this figure, $\epsilon=10^{-6}$.}
		\label{overlap_crit_fig}
	\end{center}
\end{figure*}

We plot in Figure~\ref{overlap_crit_fig} the overlap criteria~(\ref{impli_crit}) for $\eps=10^{-6}$, the two approximations~(\ref{circ_crit}) and (\ref{hecc_crit}) from \citep{Deck2013}, as well as the collision condition used in \citep{Laskar2017} approximated for $\alpha\rightarrow 1$,
\begin{equation}
	1-\alpha\simeq e_1+e_2\simeq\sqrt{\cm}.
\end{equation}
We also plot the first MMR islands in order to show the agreement between the proposed criterion and the actual intersections.
We see that for high eccentricities, and large $1-\alpha$, the system can verify the MMR overlap stability 
criterion while allowing for collision between the planets.
For small $\alpha$, the  MMR overlap criterion alone cannot account for the stability of the system.

\section{Critical AMD and MMR}
\label{sec.C_cecc}
\subsection{Critical AMD in a context of resonance overlap}

In \citep{Laskar2017}, we present the AMD-stability criterion based on the conservation of AMD.
We assume the system dynamics to be secular chaotic.
As a consequence the averaged semi-major axis and the total averaged AMD are conserved.
Moreover, in this approximation the dynamics is limited to random AMD exchanges between planets
with conservation of the total AMD.
Based on these assumptions, collisions between planets are possible only if the AMD of the system can be distributed such that
the eccentricities of the planets allow for collisions.
Particularly, for each pair of adjacent planets, there exists a critical AMD, noted $C_c(\alpha,\gamma)$, such that for smaller AMD, collisions are forbidden.

The critical AMD was determined thanks to the limit collision condition
\begin{equation}
	\alpha(1+e_1)=1-e_2.
	\label{colli_cond}
\end{equation}
However, in practice, the system may become unstable long before orbit intersections; in particular the secular assumption does not hold
if the system experiences chaos induced by MMR overlap.
We can, though, consider that if the islands do not overlap, the AMD is, on average, conserved 
on timescales of order $\eps^{-2/3}$ (\emph{i.e.}, of the order of the libration timescales).
Therefore, the conservation, on average,  of  the AMD is ensured as long as the system adheres to the above criteria for any distribution of the AMD between planets.
Based on the model of \citep{Laskar2017}, we compute a critical AMD associated to the criterion (\ref{impli_crit}).

We consider a pair as AMD-stable if no distribution of AMD between the two planets allows the overlap of MMR.
A first remark is that no pair can be considered as AMD-stable if $\alpha>\acir$, because in this case,
even the circular orbits lead to MMR overlap.
Let us write the criterion~(\ref{impli_crit}) as a function of $\alpha$ and $\eps$;
\begin{equation}
\sqrt{\cm}=g(\alpha,\eps),
\end{equation}
where
\begin{align}
	g(\alpha,\epsilon)&=\frac{3^4(1-\alpha)^5}{2^9r\epsilon}-\frac{32r\epsilon}{9(1-\alpha)^2} & \alpha<\acir,\nnb
	&=0& \alpha>\acir.
\end{align}
$\sqrt{\cm}$ depends on $\Delta\varpi$ and has a maximum for $\Delta\varpi=\pi$.
Since the variation of $\Delta\varpi$ does not affect the AMD of the system,
we fix $\Delta\varpi=\pi$ since it is the least-favorable configuration. 
Therefore we have 
\begin{equation}
\sqrt{\cm}=c_1+c_2.
\end{equation}

\begin{figure}
	\begin{center}
		\includegraphics[width=8cm]{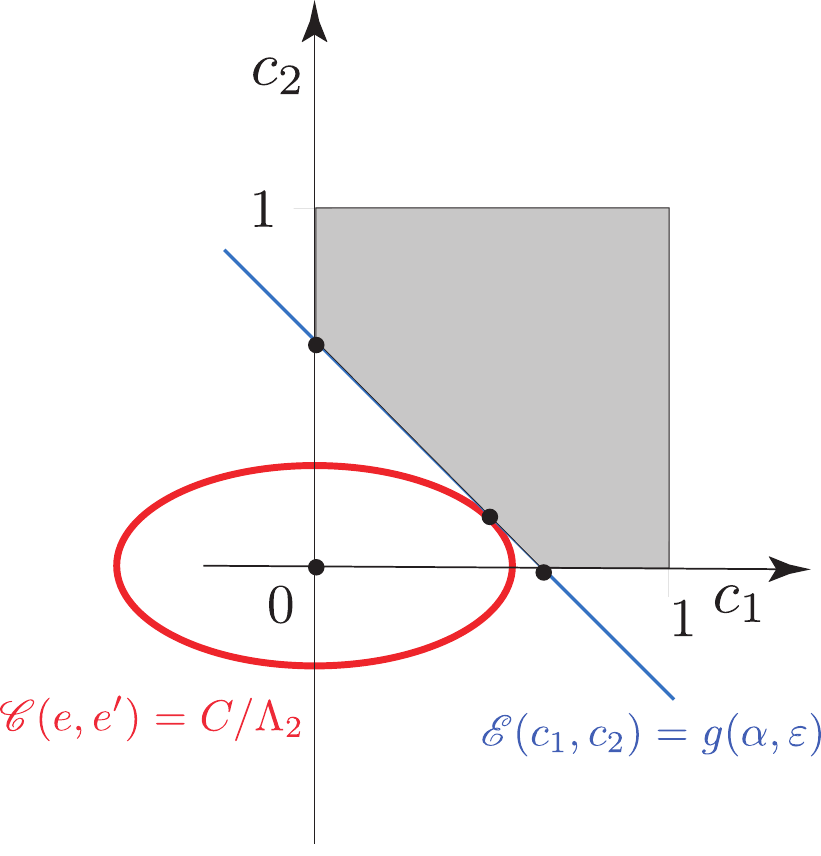}
		\caption{MMR overlap criterion represented in the $(c_1,c_2)$ plane.}
		\label{fig_schema}
	\end{center}
\end{figure}

We define the relative AMD of a pair of planets $\cC$ and express it as a function of the variables $c_i$
\begin{equation}
	\cC=\frac{C}{\Lambda_2}=\frac{1}{2}\left(\gamma\sqrt{\alpha}c_1^2+c_2^2\right).
	\label{rel_AMD}
\end{equation}
The critical AMD $\Cce$ associated to the overlap criterion (\ref{impli_crit}) can be defined as
the smallest value of relative AMD such that the conditions
\begin{align}
	\cE(c_1,c_2)&=c_1+c_2=g(\alpha,\eps)\label{rel_cE}\\
	\cC(c_1,c_2)&=\frac{1}{2}\left(\gamma\sqrt{\alpha}c_1^2+c_2^2\right)=\Cce
\end{align}
are verified by any couple $(c_1,c_2)$. We represent this configuration in Figure~\ref{fig_schema}.
As in \citep{Laskar2017}, the critical AMD is obtained through Lagrange multipliers
\begin{equation}
\nabla\cC\propto\nabla\cE.
\end{equation}
The tangency condition gives a relation between $c_1$ and $c_2$,
\begin{equation}
	\gamma\sqrt{\alpha}c_1=c_2.
\end{equation} 
Replacing $c_2$ in relation~(\ref{rel_cE}) gives the critical expression of $c_1$ and we immediately obtain the expression of $c_2$
\begin{equation}
	c_{c,1}=\frac{g(\alpha,\eps)}{1+\gamma\sqrt{\alpha}} \quad c_{c,2}=\frac{\gamma\sqrt{\alpha}g(\alpha,\eps)}{1+\gamma\sqrt{\alpha}}.
\end{equation}
The value of $\Cce$ is obtained by injecting the critical values $c_{c,1}$ and $c_{c,2}$ into the expression of $\cC$
\begin{equation}
	\Cce(\alpha,\gamma,\eps)=\frac{g(\alpha,\eps)^2}{2}\frac{\gamma\sqrt{\alpha}}{1+\gamma\sqrt{\alpha}}.
	\label{Cce}
\end{equation}

\subsection{Comparison with the collision criterion}

It is then natural to compare the critical AMD $\Cce$ to the critical AMD $C_c$ (denoted hereafter by $\Cc$)  derived from the collision condition (\ref{colli_cond}).
If $\alpha>\acir$, the circular overlap criterion implies that $\Cce=0$ and therefore $\Cce$ should be preferred to the previous criterion $\Cc$.
However, $\Cce$ was obtained thanks to the assumption that $\alpha$ was close to 1.
Particularly, it makes no sense to talk about first-order MMR overlap for $\alpha<0.63$ which corresponds to the center of the MMR 2:1.
Therefore, the collision criterion should be used for small $\alpha$. We need then to find $\alpha_R$ such that for
$\alpha<\alpha_R$, we should use the critical AMD $\Cc$.
Since we are close to 1, we use a development of $\Cc$ presented in \citep{Laskar2017}, and similarly, only keep the leading terms in $1-\alpha$ in $\Cce$. The two expressions are
\begin{equation}
	\Cc=\frac{\gamma}{1+\gamma}\frac{(1-\alpha)^2}{2}, \quad \Cce=\frac{\gamma}{1+\gamma}\frac{g(\alpha,\eps)^2}{2}.
\end{equation}
We observe that for $\alpha$ close to 1, the two expression have the same dependence on $\gamma$, therefore, $\alpha_R$ depends solely on $\epsilon.$
Simplifying $\Cc=\Cce$ gives $\alpha_R$ as a solution of the polynomial equation in $(1-\alpha)$;
\begin{equation}
	3^6(1-\alpha)^7-3^22^9r\eps(1-\alpha)^3-2^{14}(r\eps)^2=0.
	\label{eq_aR}
\end{equation}
While an exact analytical solution cannot be provided, a development in powers of $\eps$ gives the following expression
\begin{align}
	1-\alpha_R&=\frac{4}{3}(2r\eps)^{1/4}+\frac{1}{4}\sqrt{2r\eps}+\O(\eps^{3/4})\nnb
	          &=1.50\eps^{1/4}+0.316\sqrt{\eps}+\O(\eps^{3/4}).
	          \label{alphaR}
\end{align}
It should be remarked that the first term can be directly obtained using Deck's high-eccentricity approximation.

\begin{figure}
	\begin{center}
		\includegraphics[width=8cm]{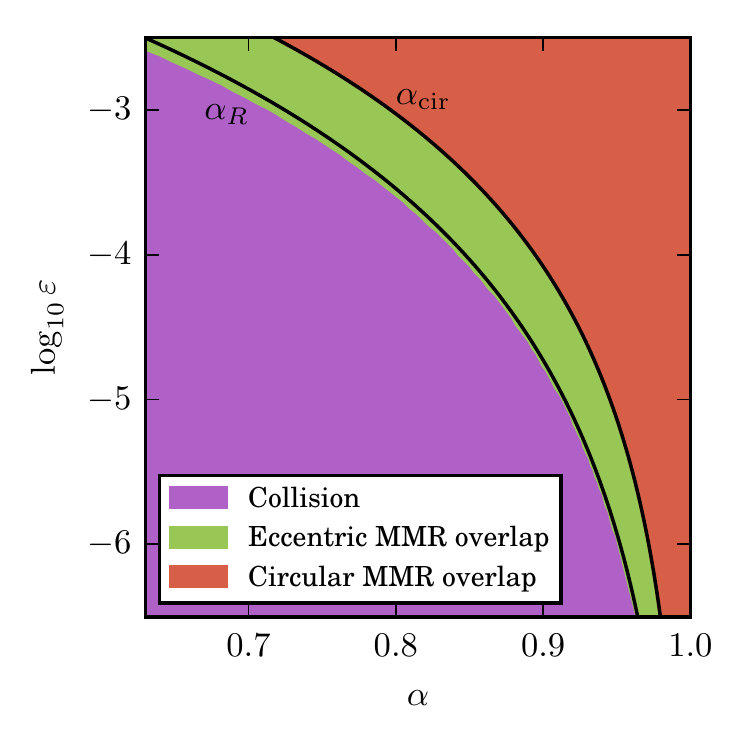}
		\caption{Regions of application of the different criteria presented in this work. The purple region  represents $\Cc$ is the smallest, in the green zone, $\Cce$ is the smallest and the circular overlap criterion is verified in the red zone. We see that the curve $\alpha_R$ computed through a development of $\Cc$ and $\Cce$ presents a good agreement with the real limit between the green and the purple area. Here $\gamma=1$.}
		\label{fig_diff_crit}
	\end{center}
\end{figure}

In Figure~\ref{fig_diff_crit} we plot $\alpha_R$ and $\acir$ and indicate which criterion is used in the areas delimited by the curves.
We specifically represented  the region $\alpha>\acir$ because we cannot treat this
region in a similar manner to the remaining region since comparing the relative AMD $\cC$ to $\Cce$ does not provide any information.
We see that the curve $\alpha_R$ is not exactly at the limit where $\Cc=\Cce$ for higher $\epsilon$ due to the development of the critical AMDs for $\alpha\rightarrow1$.
We study the influence of $\gamma$ on the difference between $\alpha_R$ and the actual limit in Appendix \ref{app_gam}

\begin{figure}
\begin{center}
\includegraphics[width=8cm]{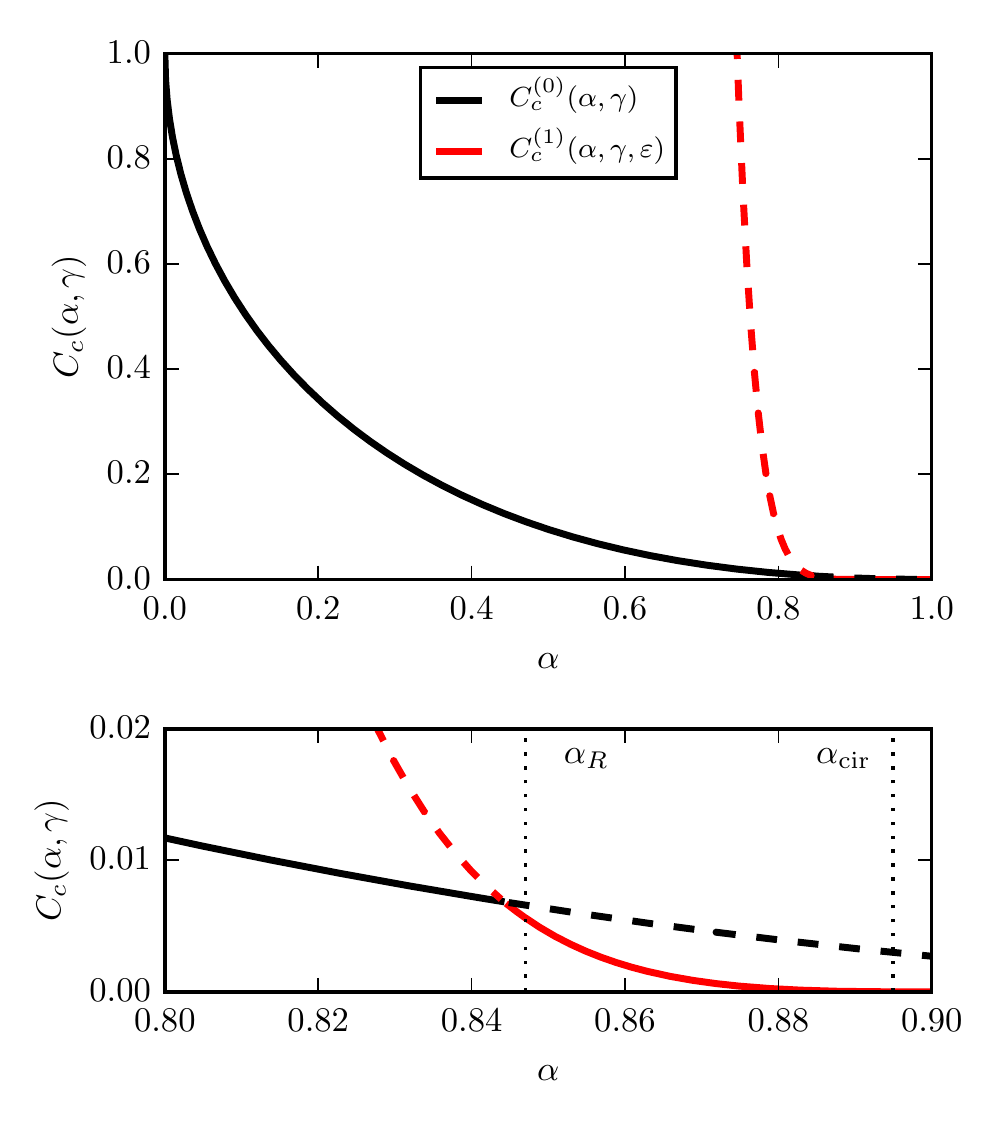}
\caption{Representation of the two critical AMD presented in this paper. $\Cc$ in black is the collisional criterion from \citep{Laskar2017}, $\Cce$ in red is the critical AMD derived from the MMR overlap criterion. In this plot,
$\eps=10^{-4}$ and $\gamma=1$.}
\label{pwCc}
\end{center}
\end{figure}

For stability analysis, we need to choose the smallest of the two critical AMD. For $\alpha<\alpha_R$,
the collisional criterion is better and the MMR overlap criterion is used for $\alpha>\alpha_R$.
We thus define a piece-wise global critical AMD represented in Figure~\ref{pwCc}
\begin{align}
	C_c(\alpha,\gamma,\eps)&=\Cc(\alpha,\gamma) &\alpha<\alpha_R(\eps,\gamma),\nnb
	                       &=\Cce(\alpha,\gamma,\epsilon) &\alpha>\alpha_R(\eps,\gamma).
	\label{amd_crit}
\end{align}

\section{Effects of the MMR overlap on the AMD-classification of planetary systems}
\label{sec_class}

In \citep{Laskar2017}, we proposed a classification of the planetary systems based on their AMD-stability.
A system is considered as AMD-stable if every adjacent pair of planets is AMD-stable.
A pair is considered as AMD-stable if its AMD-stability coefficient
\begin{equation}
\beta=\frac{C}{\Lambda'\Cc}<1,
\end{equation}
where $C$ is the total AMD of the system, $\Lambda'$ is the circular momentum of the outer planet and $\Cc$ is the critical AMD derived from the collision condition.
A similar AMD-coefficient can be defined using the global critical AMD defined in (\ref{amd_crit}) 
instead of the collisional critical AMD $\Cc$.
Let us note $\bM$, the AMD-stability coefficient associated to the critical AMD~(\ref{amd_crit}).

We can first observe that $\bM$ is not defined for $\alpha>\acir$.
Indeed, the conservation of the AMD cannot be guaranteed for orbits experiencing short-term chaos.

We use the modified definition of AMD-stability in order to test its effects on the AMD-classification proposed in \citep{Laskar2017}.

\begin{figure}[htbp]
	\begin{center}
		\includegraphics[width=8.5cm]{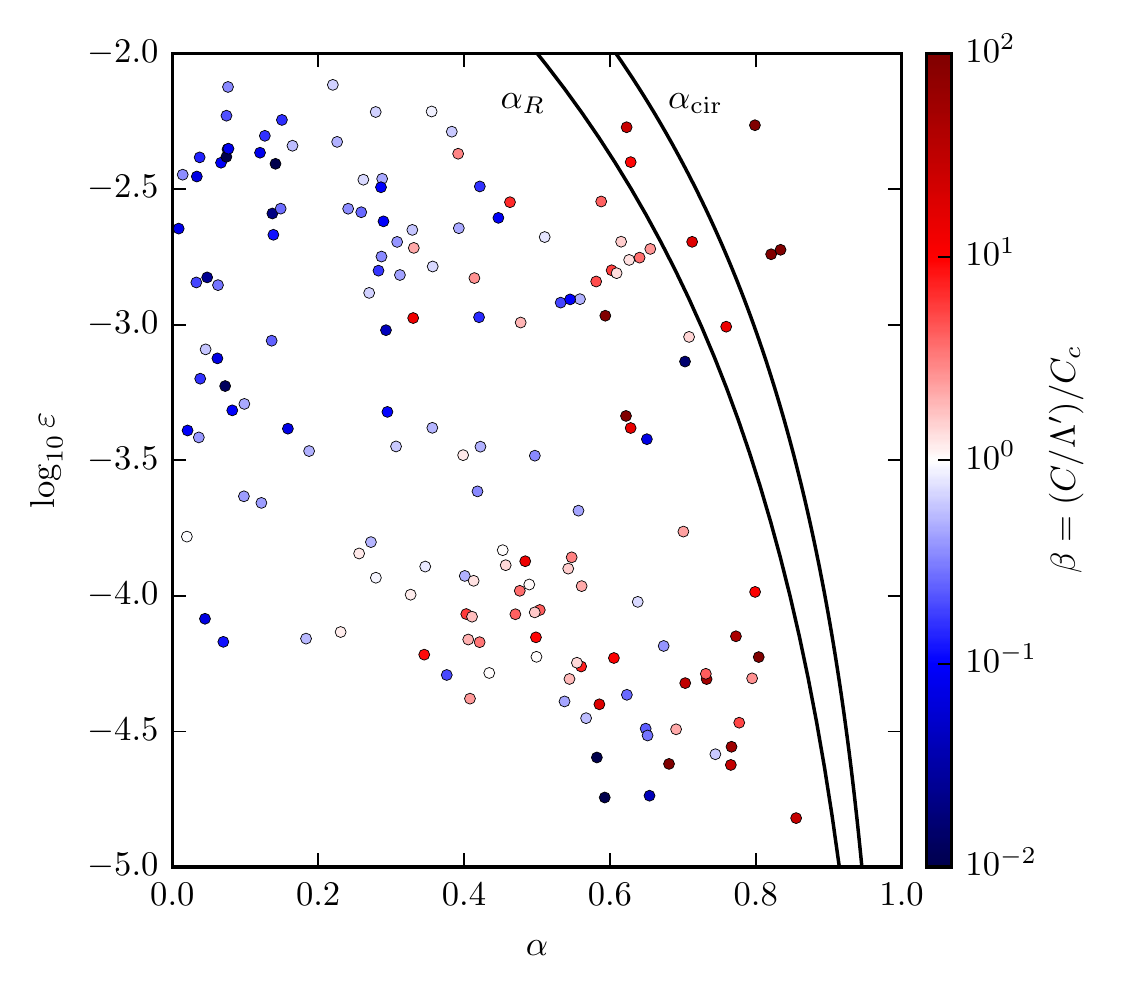}
		\caption{Pairs of adjacent planets represented in the ${\alpha-\eps}$~plane. The color corresponds to the AMD-stability coefficient. We plotted the two limits $\alpha_R$ corresponding to the limit between the collision and the MMR-overlap-based criterion and $\acir$ corresponding to the MMR overlap for circular orbits.}
		\label{aleps_ratios}
	\end{center}
\end{figure}

\subsection{Sample and methodology}
\label{methodo}
We first briefly recall the methodology used in \citep{Laskar2017}; to which we refer the reader for full details.
 We compute the AMD-stability coefficients for the systems taken from the Extrasolar Planets Encyclop\ae dia\footnote{http://exoplanet.eu/} with known periods, planet masses, eccentricities, and stellar mass.
For each pair of adjacent planets, $\eps$ was computed using the expression
\begin{equation}
	\eps=\frac{m_1+m_2}{m_0},
\end{equation}
where $m_1$ and $m_2$ are the two planet masses and $m_0$, the star mass.
The semi-major axis ratio was derived from the period ratio and Kepler third law in order to reduce the uncertainty.

The systems are assumed coplanar, however in order to take into account the contribution
of the real inclinations to the AMD, we define $C_p$, the coplanar AMD of the system, defined as the AMD of the same system if it was coplanar.
We can compute coplanar AMD-stability coefficients $\bM_p$ using $C_p$ instead of $C$, and we define the total AMD-stability coefficients as  $\beta=2\bM_p$.
Doing so, we assume the equipartition of the AMD between the different degree of freedom of the system.

We assume the uncertainties of the database quantities to be Gaussian.
For the eccentricities, we use the same method as in the previous paper.
The quantity $e\cos\varpi$ is assumed to be Gaussian with the mean, the value of the database and standard deviation, the database uncertainty. The quantity $e\sin\varpi$ is assumed to have a Gaussian distribution with zero mean
and the same standard deviation. The distribution of eccentricity is then derived from these two distributions.

We then propagate the uncertainties through the computations thanks to Monte-Carlo simulations of the original distributions.
For each of the systems, we drew 10,000 values of masses, periods and eccentricities from the computed distributions.
We then compute $\bM$ for each of these configurations and compute the 1-$\sigma$ confidence interval.

In \citep{Laskar2017}, we studied 131 systems but we did not find the stellar mass for 4 of these systems.
They were, as a consequence, excluded from this study.
Moreover, the computation of $\eps$ for the pairs of planets of the 127 remaining systems of the sample led in some cases to high planet-to-star mass ratios.
We decide to exclude the systems such that $\acir$ was smaller than the center of the resonance 2:1.
We thus discard systems such that a pair of planets has
\begin{equation}
	\eps>\eps_\mathrm{lim}=8.20\times10^{-3}.
\end{equation}
As a result, we only consider in this study 111 systems that meet the above requirements.

A pair is considered stable if the 1-$\sigma$ confidence interval (84\% of the simulated systems) of the AMD-stability coefficient $\bM$ is below 1.
A system is stable if all adjacent pairs are stable.

\subsection{Results}

\begin{figure}
	\begin{center}
		\includegraphics[width=8cm]{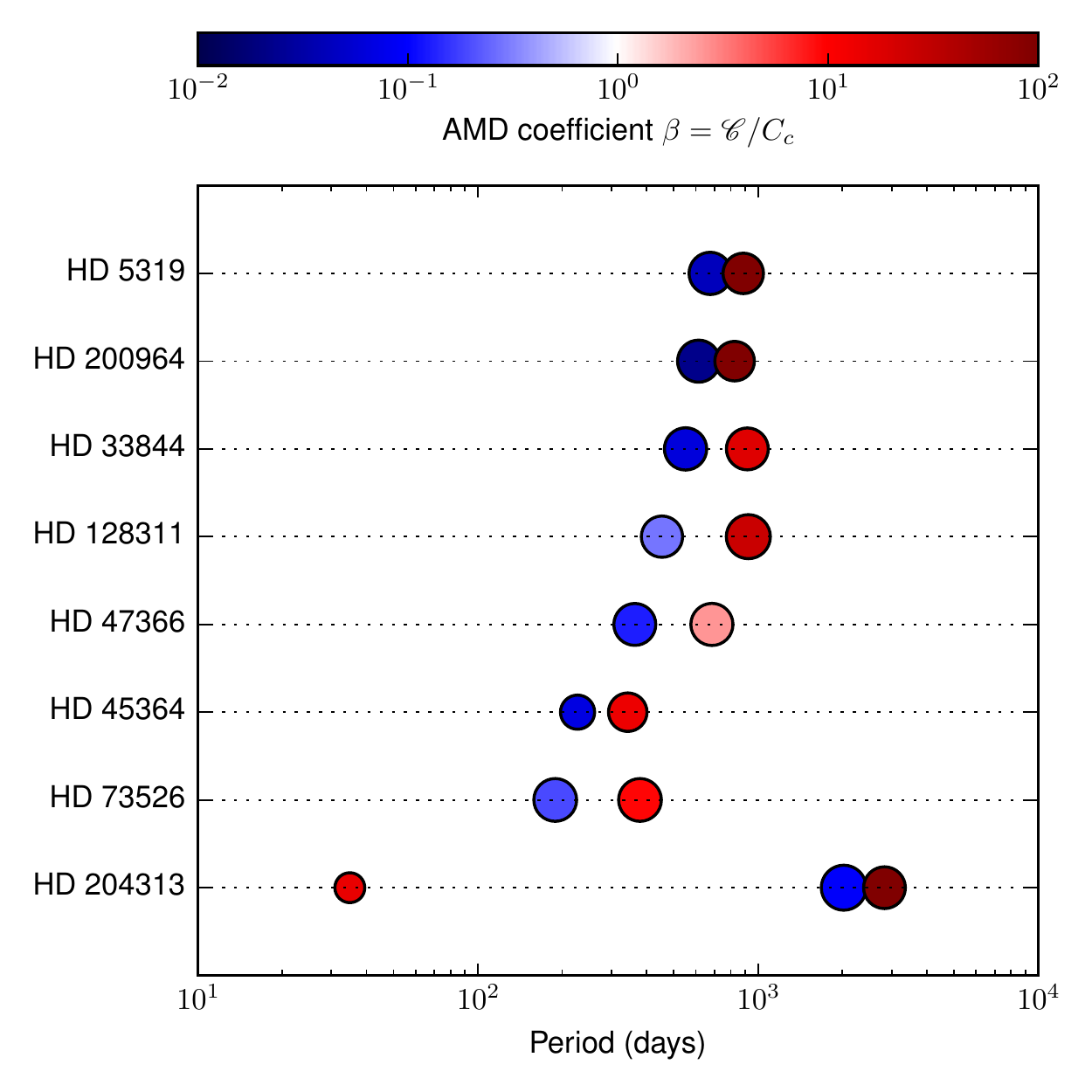}
		\caption{Architecture of the systems where the MMR overlap criterion changes the AMD-stability. The color corresponds to the value of the AMD-stability coefficient associated with the inner pair. For the innermost planet, it corresponds to the star AMD-stability criterion \citep{Laskar2017}. The diameter of the circle is proportional to the log of the mass of the planet.}
		\label{archi}
	\end{center}
\end{figure}

\begin{figure}
	\begin{center}
		\includegraphics[width=8cm]{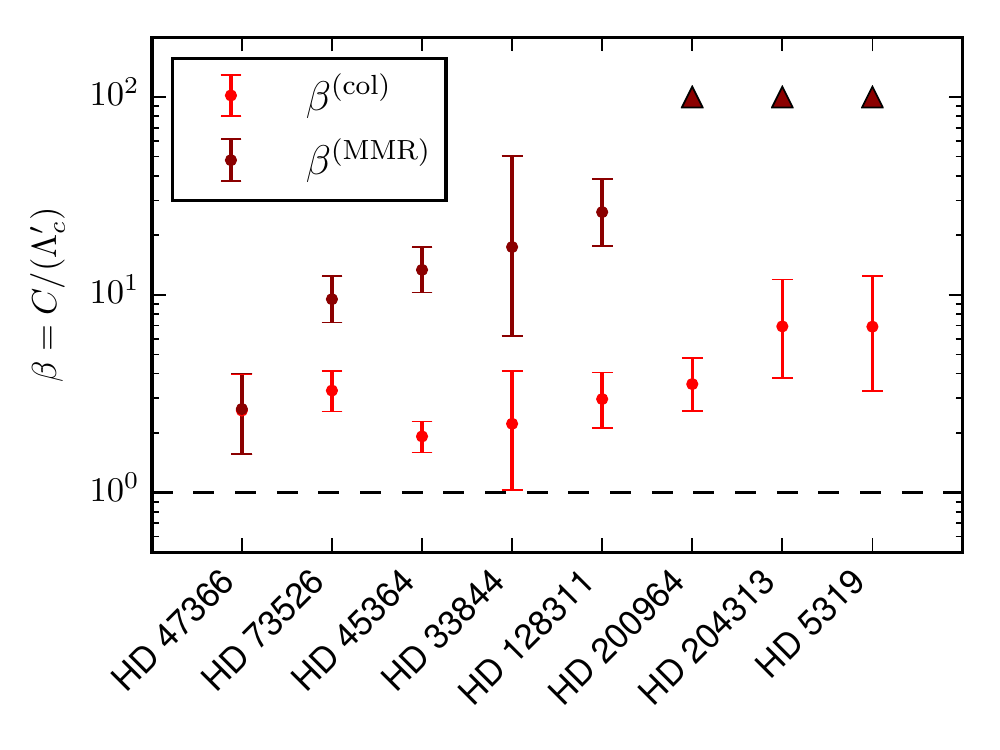}
		\caption{AMD-stability coefficient of the pairs affected by the change of criterion. $\beta^{\mathrm{(col)}}$ corresponds to the coefficient computed with the collisional critical AMD, and $\bM$  refers to the one computed with the MMR overlap critical AMD. The triangles represent the pairs where $\bM$ goes to infinity.}
		\label{change}
	\end{center}
\end{figure}

Figure~\ref{aleps_ratios} shows the planet pairs of the considered systems  in a plane $\alpha$-$\eps$.
The color associated to each point is the AMD-stability coefficient of the pair.
The values chosen for the plot correspond for all quantities to the median.
We remark that very few systems are concerned by the change of the critical AMD, indeed, only eight systems\footnote{
	It should be noted that for one of the systems, the MMR overlap criterion was preferred in 16\% of the Monte Carlo simulations.} have a pair of planets such that $\Cce<\Cc$.
The 111 considered systems contain 162 planet pairs plotted in Figure~\ref{aleps_ratios}.
This means that less than 5\% of the pairs are in a configuration leading to MMR overlap.

We plot in Figure~\ref{archi}, the architecture of these eight systems and give in Table \ref{table_betas} the values of the AMD-stability coefficients.
For each of these systems, the pair verifying the MMR overlap criterion was already considered AMD-unstable by the criterion based on the collision.

In order to show this, we plot in Figure~\ref{change} the AMD-stability coefficients computed with both critical AMD.
We see that the pairs affected by the change of criterion were already considered AMD-unstable in the purely secular dynamics.
However, while those pairs have a collisional AMD-coefficient $\beta$ between 1 and 10,
the global AMD-stability coefficient is increased by roughly an order of magnitude for the four pairs with $\alpha$ between $\alpha_R$ and $\acir$.
The AMD-coefficient is not defined for the three pairs verifying the circular MMR overlap criterion.
The pair HD 47366 b/c does not see a significant change of its AMD-stability coefficient due to the small number of cases where $\Cc>\Cce$.

We identify three systems, HD 200964, HD204313 and HD 5319, that satisfy the circular overlapping criterion.
As already explained in \citep{Laskar2017}, AMD-unstable planetary systems may not be dynamically unstable.
However, it should be noted that the period ratios of the AMD-unstable planet pairs are very close to particular MMR.

Indeed, we have
\begin{align}
\frac{T^{\mathrm{HD\ 200964}}_{\rm c}}{T^{\mathrm{HD\ 200964}}_{\rm b}}=1.344 \simeq 4/3,\\
\frac{T^{\mathrm{HD\ 204313}}_{\rm d}}{T^{\mathrm{HD\ 204313}}_{\rm c}}=1.399 \simeq 7/5,\\
\frac{T^{\mathrm{HD\ 5319}}_{\rm c}}{T^{\mathrm{HD\ 5319}}_{\rm b}}=1.313 \simeq 4/3.
\end{align}
The AMD-instability of those systems strongly suggests that they are indeed into a resonance which stabilizes their dynamics.

\section{Conclusions}

As shown in \cite{Laskar2017}, the notion of AMD-stability is a powerful tool to characterize the stability of planetary systems.
In this framework, the dynamics of a system is reduced to the AMD transfers allowed by the secular evolution.

However, we need to ensure that the system dynamics can be averaged over its mean motions.
While a system can remain stable and the AMD or semi-major axis can be averaged over timescales longer than the 
libration period in presence of MMR, the system stability and particularly the conservation of the AMD is 
no longer guaranteed if the system experiences MMR overlap.
In this paper, we use the MMR overlap criterion as a condition to delimit the zone of the phase space where the dynamics can be considered as secular.

We refine the criteria proposed by \citep{Wisdom1980,Mustill2012,Deck2013} and demonstrate that it is possible 
to obtain a global expression (\ref{impli_crit}), valid for all cases. The previous circular (\ref{circ_crit}) and eccentric (\ref{hecc_crit}) criteria an then be derived from (\ref{impli_crit}) as particular approximations.
Moreover, we show that expression (\ref{impli_crit}) can be used to directly  take 
into account the first-order MMR in the notion of AMD-stability.

With this work on first-order MMR, we improve the AMD-stability definition by
addressing the problem of the minimal distance between close orbits. For
semi-major axis ratios $\alpha$ above a given threshold $\alpha_\mathrm{cir}$
(\ref{circ_crit}), that is, $\alpha_\mathrm{cir} < \alpha < 1$, the system is considered
unstable whichever value the AMD may take given that even two circular orbits
satisfy the MMR overlap criterion. At wider separations, circular orbits are
stable but as eccentricities increase two outcomes may happen: Either the
system enters a region of MMR overlap or the collision condition is
reached. The system is said to be AMD-unstable as soon as any of these
conditions is reached. Above a second threshold, $\alpha_R < \alpha <
\alpha_\mathrm{cir}$ (Eq.~\ref{alphaR}) the AMD-stability is governed by MMR overlap
while for wider separations ($\alpha < \alpha_R$) we retrieve the critical AMD
defined in \citep{Laskar2017} which only depends on the collision
condition.

In order to  improve the AMD-stability definition for the collision region, we could even take into account the non-secular dynamics induced by higher-order MMR and close-encounter consequences on the AMD.
To study this requires more elaborated analytical considerations than those presented here that are restricted to the first-order MMR; this will be the goal of future work.

We show in Section \ref{sec_class} that very few systems satisfy the circular MMR overlap criterion.
Moreover, the presence of systems satisfying this criterion strongly suggests that they are protected by a particular MMR. In this case, the AMD-instability is a simple tool suggesting unobvious dynamical properties.

\bibliographystyle{aa}
\bibliography{MMRAMD}

\appendix

\section{Expression of the first-order resonant Hamiltonian}
\label{res_ham_appendix}

We use the method proposed in \citep{Laskar1991} and \citep{Laskar1995} to determine the expression of the planetary perturbation $\hH_1$.
$\hH_1$ can be decomposed into a part from the gravitational potential between planets $\hU_1$ and a kinetic part $\hT_1$ as
\begin{equation}
\epsilon\hH_1=\hU_1+\hT_1,
\end{equation}
with
\begin{align}
\hU_1&=-\G \frac{m_1m_2}{\Delta_{12}}=-\frac{m_1}{m_0}\frac{\mu^2m_2^3}{\Lambda_2^2}\frac{a_2}{\Delta_{12}}\\
\hT_1&=\frac{\tu_1\cdot\tu_2}{m_0}+\frac{1}{2m_0}(\|\tu_1\|^2+\|\tu_2\|^2).
\end{align}

The difficulty comes from the development of $a_2/\Delta_{12}$ and its expression in terms of Poincar\'e variables.
We note $S$, the angle between $\u_1$ and $\u_2$. We have
\begin{equation}
\Delta_{12}^2=u_1^2+u_2^2-2u_1u_2\cos S.
\end{equation}
Let us denote $\rho=u_1/u_2$, $a_2/\Delta_{12}$ can be rewritten
\begin{align}
\frac{a_2}{\Delta_{12}}&=\frac{a_2}{u_2}\left(1+\rho^2-2\rho\cos S\right)^{-1/2}\nnb
&=\frac{a_2}{u_2}\left(A+V\right)^{-1/2},
\label{asD}
\end{align}
where we denote
\begin{align}
A&=1+\alpha^2-2\alpha\cos(\lambda_1-\lambda_2),\\
V&=\alpha^2V_2+2\alpha V_1,\\
V_1&=\cos(\lambda_1-\lambda_2)-\frac{\rho}{\alpha}\cos S,\\
V_2&=\left(\frac{\rho}{\alpha}\right)^2-1.
\end{align}
$V$ is at least of order one in eccentricity. We can therefore develop (\ref{asD}) for small $V$.
We only keep the terms of first order in eccentricity,
\begin{equation}
\frac{a_2}{\Delta_{12}}=\frac{a_2}{u_2}A^{-1/2}-\frac{1}{2}\frac{a_2}{u_2}VA^{-3/2} +\O(V^2).
\label{asD.dev}
\end{equation}

The well-known development of the circular coplanar motion $A$ gives \citep[e.g.,][]{Poincare1905}
\begin{equation}
A^{-s}=\frac{1}{2}\sum_{k\in\Z} \lc{s}{k}{\alpha}\expo{\i k(\lambda_1-\lambda_2)},
\end{equation}
where $\lc{s}{k}{\alpha}$ are the Laplace coefficients (\ref{Lapcoef}).

Because of the averaging over the non-resonant fast angles, the non-vanishing terms have a dependence on $\lambda_i$ of the form $j\left((p+1)\lambda_2-p\lambda_1\right)$.
Since we only keep  the terms of first order in eccentricity, the d'Alembert's rule (\ref{dalemb}) imposes $j=\pm 1$.
Let us compute the first-order development of $a_2/u_2$ and $V$ in terms of Poincar\'e variables and combine these expressions with $A^{-1/2}$ and $A^{-3/2}$ in order to select the non-vanishing terms.

Let us denote $z_i=\e^{\i\lambda_i}$ and $z=z_1\bar{z}_2=\e^{\i(\lambda_1-\lambda_2)}$. The researched terms are of the form
\begin{align}
\expo{\i((p+1)\lambda_2-p\lambda_1)}&=z_2z^{-p}=z_1z^{-(p+1)}\\
\expo{-\i((p+1)\lambda_2-p\lambda_1)}&=\bar z_2z^{p}=\bar z_1z^{p+1}.
\end{align}
Let us denote 
\begin{equation}
X_i=\hx_i\sqrt{\frac{2}{\hL_i}}=\sqrt{\frac{2\hC_i}{\hL_i}}\expo{-\i\varpi_i}=e_i \expo{-\i\varpi_i}+\O(e_i^2),
\end{equation}
 the first term in the development~(\ref{asD.dev}) gives 
\begin{equation}
\frac{a_2}{u_2}A^{-1/2}=\frac{1}{2}\left(1+\frac{1}{2}X_2z_2+\frac{1}{2}\bar X_2\bar  z_2\right)\sum_{k\in\Z}\lc{1/2}{k}{\alpha}z^k +\O(e_2^2).
\end{equation}
The contributing term has for expression
\begin{equation}
\frac{1}{4}\lc{1/2}{p}{\alpha}(\X_2+\bar{\X_2}),
\label{r_21}
\end{equation}
where $\X_i=\hat \x_i\sqrt{2/\hL_i}=X_i \expo{\i((p+1)\lambda_2-p\lambda_1)}$.

For the computation of the second term of (\ref{asD.dev}), the only contribution comes from $V$ since $a_2/u_2\sim1$. We define
\begin{align}
U&=X_1z_1-X_2z_2\nnb
&=\sqrt{\frac{2\hC_1}{\hL_1}}\expo{\i(\lambda_1-\varpi_1)}-\sqrt{\frac{2\hC_2}{\hL_2}}\expo{\i(\lambda_2-\varpi_2)}.
\end{align}
$V$ can be expressed as a function  of $z,\bar z, U$ and $\bar U$. Indeed we have
\begin{equation}
\frac{\rho}{\alpha}= 1-\frac{1}{2}(U+\bar U)+\O(e^2)
\end{equation}
and
\begin{equation}
\cos S=\frac{1}{2}\left(z+\bar z+U(z-\bar z)+\bar U (\bar z -z)\right)+\O(e^2),
\end{equation}
where $\O(e^2)$ corresponds to terms of total degree in eccentricities of at least 2. We deduce from these two last expressions that
\begin{align}
V_1&=\frac{1}{4}\left(U(3\bar z-z)+\bar U (3z-\bar z)\right)+\O(e^2),\\
V_2&=-(U+\bar U)+\O(e^2).
\end{align}
We can therefore write\footnote{
	In \cite{Laskar1995} the first-order expression of $V$ is written $W_1=(UZ+\bar U\bar Z)$ instead of ${W_1=(UZ+\bar U\bar Z)/2}$. This misprint in equation (47) of \citep{Laskar1995}  is transmitted as well in equation (51). It has no consequences in the results of the paper.} 
\begin{equation}
V=\frac{1}{2}(UZ+\bar U\bar Z)+\O(e^2),
\end{equation}
where $Z=\alpha(3\bar z -2\alpha -z)$.
With this expression of $V$, it is easy to gather the corresponding terms and the second term in the development~(\ref{asD.dev}) gives the contributing term
\begin{align}
-&\frac{\alpha}{8}\left(3\lc{3/2}{p}{\alpha}-2\alpha\lc{3/2}{p+1}{\alpha}-\lc{3/2}{p+2}{\alpha}\right)\left(\X_1+\bar\X_1\right)+\nnb
&\frac{\alpha}{8}\left(3\lc{3/2}{p-1}{\alpha}-2\alpha\lc{3/2}{p}{\alpha}-\lc{3/2}{p+1}{\alpha}\right)\left(\X_2+\bar\X_2\right)
\label{r_3}.
\end{align}

After gathering the terms~(\ref{r_21},\ref{r_3}), we can give the expression of the resonant Hamiltonian
\begin{equation}
\hH=\hK+\hat R_1(\hat \x_1+\bar{\hat \x}_1)+\hat R_2(\hat\x_2+\bar{\hat\x}_2),
\end{equation}
where
\begin{align}
\hat R_1&= -\epsilon\frac{\gamma}{1+\gamma}\frac{\mu^2m_2^3}{\hL_2^2}\frac{1}{2}\sqrt{\frac{2}{\hL_1}}r_1(\alpha),\\
\hat R_2&= -\epsilon\frac{\gamma}{1+\gamma}\frac{\mu^2m_2^3}{\hL_2^2}\frac{1}{2}\sqrt{\frac{2}{\hL_2}}r_2(\alpha)\\
\end{align}
with $\gamma=m_1/m_2$, and
\begin{align}
r_1(\alpha)&=-\frac{\alpha}{4}\left(3\lc{3/2}{p}{\alpha}-2\alpha\lc{3/2}{p+1}{\alpha}-\lc{3/2}{p+2}{\alpha}\right),
\label{coef_r1.app} \\
r_2(\alpha)&=\frac{\alpha}{4}\left(3\lc{3/2}{p-1}{\alpha}-2\alpha\lc{3/2}{p}{\alpha}-\lc{3/2}{p+1}{\alpha}\right)\nnb
&\quad +\frac{1}{2}\lc{1/2}{p}{\alpha}.
\label{coef_r2.app}
\end{align}

The kinetic part $\hT_1$ has no contribution to the averaged resonant Hamiltonian for $p>1$.
Indeed, as explained above, due to the d'Alembert rule, the first-order terms must have an angular dependence 
of the form $j(-p\lambda_1+(p+1)\lambda_2)$.
At the first order in $\eps$, such a term can only be present in the development of the inner product 
$\tu_1\cdot\tu_2$.
At the first order in eccentricities, we have \citep{Laskar1995}
\begin{equation}
\tu_1\cdot\tu_2=\frac{\mu^2 m_1^2 m_2^2}{\hL_1\hL_2}\Re((\expo{\i \omega_1}+ X_1)(\expo{-\i \omega_2}+\bar{X_2})) +\O(e^2),
\end{equation}
where $\omega_j$ is the true longitude of the planet $j$.
The only term with the good angular dependence comes from $\Re \expo{\i(\omega_1-\omega_2)}$ since the other first-order terms  only  depend on one mean longitude.
The development of $\expo{\i(\omega_1-\omega_2)}$ at the first order in eccentricities gives
\begin{equation}
	\expo{\i(\omega_1-\omega_2)}= z+z_1z\bar X_1-\bar z_2X_1+z\bar z_2 X_2-z_1\bar X_2+\O(e^2).
\end{equation}
Thus for $p>1$, $\hT_1$ has no contribution to the averaged Hamiltonian, and for $p=1$ we have
\be
\H_{1,i}=\frac{1}{2m_0}\frac{\mu m_1^2}{\hL_1}\frac{\mu m_2^2}{\hL_2}(\X_2+\bar{\X}_2).
\ee

\subsection{Asymptotic expression of the resonant coefficients}
\label{Res_coeff}
We present the method we used to obtain the analytic development of the coefficients $r_1$ and $r_2$
defined in equations (\ref{coef_r1.app}) and (\ref{coef_r2.app}).
Using the expression of $\lc{s}{k}{\alpha}$, we have
\begin{align}
r_1(\alpha)=-\frac{\alpha}{4\pi}&\left[\int_{-\pi}^{\pi}\frac{3\cos(p\phi)}{(1+\alpha^2-2\alpha\cos\phi)^{3/2}}\d\phi+\right.\nnb
            &\left.\int_{-\pi}^{\pi}\frac{-2\alpha\cos((p+1)\phi)}{(1+\alpha^2-2\alpha\cos\phi)^{3/2}}\d\phi+\right.\nnb
            &\left.\int_{-\pi}^{\pi}\frac{-\cos((p+2)\phi)}{(1+\alpha^2-2\alpha\cos\phi)^{3/2}}\d\phi \right].                    
\end{align}
We can rewrite this expression
\begin{align}
r_1(\alpha)=-\frac{\alpha}{2\pi}&\left[\int_{-\pi}^{\pi}\frac{(\cos(\phi)-\alpha)\cos((p+1)\phi)}{(1+\alpha^2-2\alpha\cos\phi)^{3/2}}\d\phi+\right.\nnb
&\left.\int_{-\pi}^{\pi}\frac{2\sin\phi\sin((p+1)\phi)}{(1+\alpha^2-2\alpha\cos\phi)^{3/2}}\d\phi \right]. 
\end{align}
We make the change of variable $\phi=(1-\alpha)u$ in the integrals. Factoring $(1-\alpha)^{3}$, the denominators in the integrals can be developed for $\alpha\rightarrow1$
\begin{align}
      (1+\alpha^2-2\alpha\cos\phi)^{3/2}&=\left(1+2\alpha\frac{1-\cos((1-\alpha)u)}{(1-\alpha)^2}\right)^{3/2}\nnb
                                      &\simeq (1-\alpha)^3(1+u^2)^{3/2}.
\end{align}
Using the relation $\alpha_0=(p/(p+1))^{2/3}$, the numerators can be developed
\begin{align}
	\mathcal{N}_1&=(\cos((1-\alpha)u)-\alpha)\cos((p+1)(1-\alpha)u)\nnb
	              &\simeq(1-\alpha)\cos\left(\frac{2u}{3}\right)\\
	\mathcal{N}_2&=2\sin((1-\alpha)u)\sin((p+1)(1-\alpha)u)\nnb
	&\simeq2(1-\alpha)u\sin\left(\frac{2u}{3}\right).
\end{align}

Therefore, we deduce the equivalent of $r_1$ for $p\rightarrow+\infty$
\begin{align}
r_1(\alpha)&\sim-\frac{3(p+1)}{4\pi}\int_{-\infty}^{+\infty}\frac{\cos\left(\frac{2u}{3}\right)+2u\sin\left(\frac{2u}{3}\right)}{(1+u^2)^{3/2}}\d u\nnb
&\sim-\frac{K_1(2/3)+2K_0(2/3)}{\pi}(p+1)\\
&\sim 0.802(p+1),
\end{align}
where $K_\nu(x)$ is the modified Bessel function of the second kind.
Similarly, we have  $r_2\sim -r_1$ since the additional term is of lower order in $p$.

We can obtain the constant term of the development by using the second order expression of $\alpha_0$ and developing the integrand to the next order in $(1-\alpha)$.
We give here the numerical expressions of the two developments
\begin{align}
	r_1(\alpha_0)&=-0.802(p+1)-0.199+\O(p^{-1}),\\
	r_2(\alpha_0)&=0.802(p+1)+0.421+\O(p^{-1}).
\end{align}

\section{Development of the Keplerian part}
\label{dev_kep}

We show here that the first order in $(C-\DG)$ of the Keplerian part vanishes and give the details of the computation for the second order.
The Keplerian part can be written
\begin{align}
	\hK=&-\frac{\mu^2m_1^3}{2(\Loz-p(C-\DG))^2}\nnb
	    &-\frac{\mu^2m_2^3}{2(\Ltz+(p+1)(C-\DG))^2}.
\end{align}
Therefore, the first order in $C-\DG$ has for expression
\begin{equation}
\K_1=-\frac{\mu^2m_2^3}{\Ltz^3}\left(\frac{p\gamma^3\Ltz^3}{\Loz^3}-(p+1)\right)(C-\DG)=0,
\end{equation}
since we have
\begin{equation}
	\left(\frac{\Loz}{\Ltz}\right)^3=\gamma^3\frac{p}{p+1}.
\end{equation}
The second-order term has for coefficient
\begin{align}
\frac{1}{2}\K_2&=-\frac{3}{2} \mu^2m_2^3\left(\frac{\gamma^3p^2}{\Loz^4}+\frac{(p+1)^2}{\Ltz^4}\right)\nnb
&=-\frac{3}{2}\mu^2m_2^3(\gamma+\alpha_0)^4\left(\frac{p^2}{\gamma\left(\frac{p}{p+1}\right)^4}+\frac{(p+1)^2}{\alpha_0^4}\right)\nnb
&=-\frac{3}{2}\mu^2m_2^3(\gamma+\alpha_0)^4(p+1)^2\frac{\alpha_0^4\left(\frac{p+1}{p}\right)^2+\gamma}{\gamma \alpha_0^4}\nnb
\frac{1}{2}\K_2&=-\frac{3}{2}\mu^2m_2^3\frac{(\gamma+\alpha_0)^5}{\gamma \alpha_0^4}(p+1)^2.
\end{align}

\section{Width of the resonance island}
\label{a.width}

We detail in this Appendix the computation of the resonance island's width  \citep[see also][Appendix C]{Ferraz-Mello2007}.

\subsection{Coefficients-roots relations}
\label{coef_poly}
We first explain how the width of the resonance can be related to the position of the saddle point on the $X$-axis.
The resonant island has a maximal width on the $X$-axis.
Therefore we need to compute the expression of the intersections of the separatrices with the $X$-axis.

Let us note $\H_3$, the energy at the saddle point $(X_3,0)$.
Since the energy of the separatrices is $\H_3$ as well, the two intersections of the separatrices with the $X$-axis are the solution of the equation

\begin{equation}
\H_A(X,0)=-\frac{X^4}{8}+\frac{\I_0X^2}{2}-X=\H_3+\frac{\I_0^2}{2} =\tilde{\H}_3.
\end{equation}

This equation has three solutions $X_1^*,X_2^*$, and $X_3$ which has a multiplicity of 2.
We can therefore rewrite the equation as
\begin{equation}
(X-X_1^*)(X-X_2^*)(X-X_3)^2=X^4-4\I_0X^2+8X+8\tilde\H_3.
\label{poly_X3width}
\end{equation}

We detail here the relations between the coefficients and the roots of the polynomial equation (\ref{poly_X3width}).
We have
\begin{align}
X_1^*+X_2^*+2X_3&=0
\label{rel1}\\
X_1^*X_2^*+2X_3(X_1^*+X_2^*)+X_3^2&=-2\I_0
\label{rel2}\\
X_1^*X_2^*X_3^2=8\tilde\H_3&=-X_3^4+2\I_0X_3^2-8X_3.
\label{rel3}
\end{align}
From relation \ref{rel1}, we have directly $X_1^*+X_2^*=-2X_3$, and since
\begin{equation}
	4X_1^*X_2^*=(X_1^*+X_2^*)^2-(X_1^*-X_2^*)^2,
\end{equation}
we can express $(X_1^*-X_2^*)^2$ as a function  of $X_3$ thanks to the relations (\ref{rel2})
and (\ref{rel3})
\begin{equation}
	|X_1^*-X_2^*|=\frac{4}{\sqrt{X_3}}.
\end{equation}

We thus deduce the expressions of $X_1^*$ and $X_2^*$ as functions of $X_3$
\begin{align}
X_1^*&=-X_3-\frac{2}{\sqrt{X_3}},\\
X_2^*&=-X_3+\frac{2}{\sqrt{X_3}}.
\end{align}

As explained in section \ref{sec.widthsma}, we obtain the width of the resonance in terms of variation of $\alpha$ as a function  of $X_3$ (equation (\ref{width})).
We can use this expression to obtain the width of the resonance for particular cases detailed in the following subsections.

\subsection{Width for initially circular orbits}

In the case of initially circular orbits, the minimal AMD to enter the resonance is 0.
For $\Cm=0$, the equation~(\ref{impliX3}) gives $X_3=2^{2/3}$ as a  solution and we have
\begin{align}
\frac{\delta\alpha}{\alpha_0}&=\frac{8\times2^{1/3}r^{2/3}}{3^{2/3}}\eps^{2/3}(p+1)^{1/3}\nnb
&=4.18\ \eps^{2/3}(p+1)^{1/3}.
\label{width_circ}
\end{align}
We find here the same width of resonance as \citep{Deck2013}.

\subsection{Width for highly eccentric orbits}

If we consider a system with $\Cm\gg\chi^{2/3}$, our formalism gives us the result first proposed by
\cite{Mustill2012} and improved by \cite{Deck2013} for eccentric orbits.
In this case, we can inject the approximation~(\ref{impX3_highecc}) of $X_3$ in the expression~(\ref{width}) of $\da$ and obtain
\begin{align}
\frac{\delta\alpha}{\alpha_0}&=\frac{8\sqrt{r}}{\sqrt{3}}\sqrt{\eps(p+1)}\cm^{1/4}\\
                            & = 4.14\sqrt{\eps(p+1)}\cm^{1/4}.
\label{width_hecc}
\end{align}
This result is also similar to Deck's one, using $\sqrt{\cm}$ instead of $\sigma$ \citep[equation~(25)]{Deck2013}.

\subsection{Width for low eccentric orbits}

For $\Cm\ll \chi^{2/3}$, we propose here a new expression of the width of resonance thanks to the expression~(\ref{impX3_lowecc}).
This expression is an extension of the circular result presented above (\ref{width_circ}).
Let us develop $\sqrt{X_3}$ for $\Cm\ll \chi^{2/3}$
\begin{align}
\sqrt{X_3}&=\sqrt{2^{2/3}+\frac{2}{3^{2/3}r^{1/3}}\frac{(p+1)^{1/3}}{\eps^{1/3}}\sqrt{\cm}}\nnb
&\simeq2^{1/3}\left(1+\frac{1}{6^{2/3}r^{1/3}}\frac{(p+1)^{1/3}}{\eps^{1/3}}\sqrt{\cm}\right).
\end{align}
Therefore for low-eccentricity systems, we have
\begin{align}
\frac{\delta\alpha}{\alpha_0}&\simeq\frac{\da_c}{\alpha_0}\left(1+\frac{1}{6^{2/3}r^{1/3}}\frac{(p+1)^{1/3}}{\eps^{1/3}}\sqrt{\cm}\right),
\label{width_lecc}
\end{align}
where $\da_c$ is the width of the resonance for initially circular orbits defined in (\ref{width_circ}).

\section{Influence of $\gamma$ on the limit $\alpha_R$}
\label{app_gam}

\begin{figure}[htbp]
	\begin{center}
		\includegraphics[width=8cm]{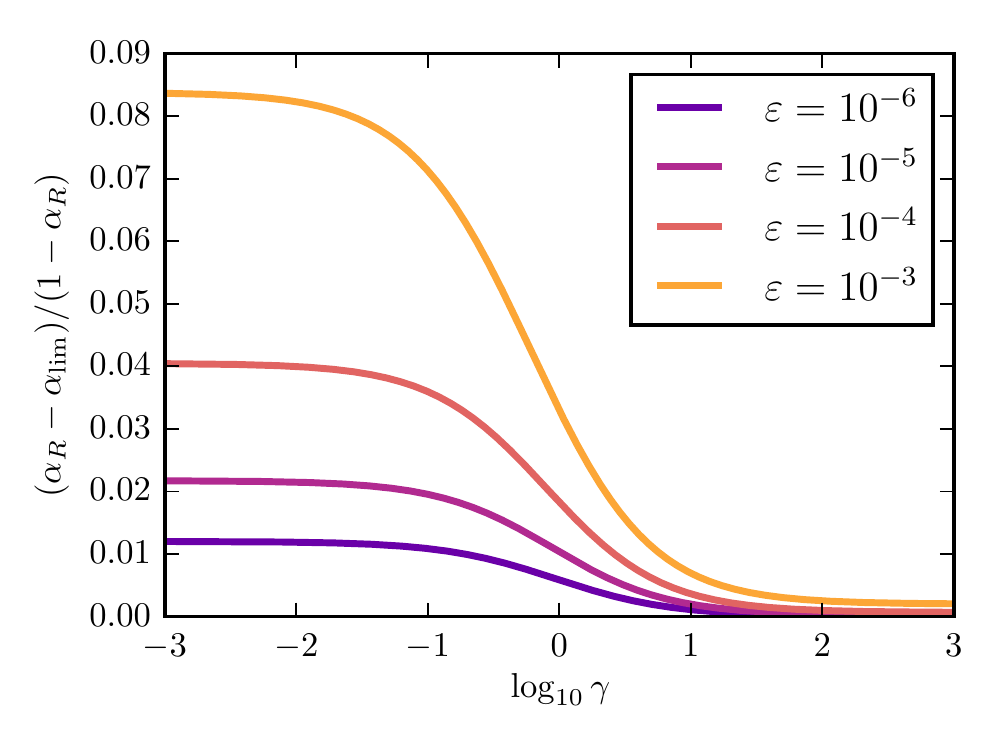}
		\caption{Difference between the limit $\alpha_\mathrm{lim}$ where $\Cc$ and $\Cce$ are equal and its approximation $\alpha_R$ scaled by $1-\alpha_R$ versus $\gamma$ for various values of $\eps$.}
		\label{delta_ar}
	\end{center}
\end{figure}

As can be seen in Figure~\ref{fig_diff_crit}, the solution $\alpha_R$ of equation (\ref{eq_aR}) is not
the exact limit where the collision and the MMR criteria are equal.
Indeed, equation  (\ref{eq_aR}) is obtained after the development of $\Cc$ and $\Cce$ for $\alpha$ close to 1.
Since at first order, both expressions have the same dependence on $\gamma$, $\alpha_R$ does not depend on $\gamma$.
In order to study the dependence on $\gamma$ of the limit $\alpha_\mathrm{lim}$ where $\Cc=\Cce$, we plot in Figure (\ref{delta_ar}), for different values of $\eps$, the quantity
\begin{equation}
\delta\alpha_R(\eps,\gamma)=\frac{\alpha_R(\eps)-\alpha_\mathrm{lim}(\eps,\gamma)}{1-\alpha_R(\eps)},
\end{equation}
which gives the error made when approximating $\alpha_\mathrm{lim}$ by $\alpha_R$.
We see that all the curves have the same shape with an amplitude increasing with $\epsilon$.
For high $\gamma$, $\alpha_R$ is very accurate even for the greatest values of $\epsilon$.
Moreover, the error is maximum for very small $\gamma$ and always within a few percent.

\begin{figure}[htbp]
	\begin{center}
		\includegraphics[width=8cm]{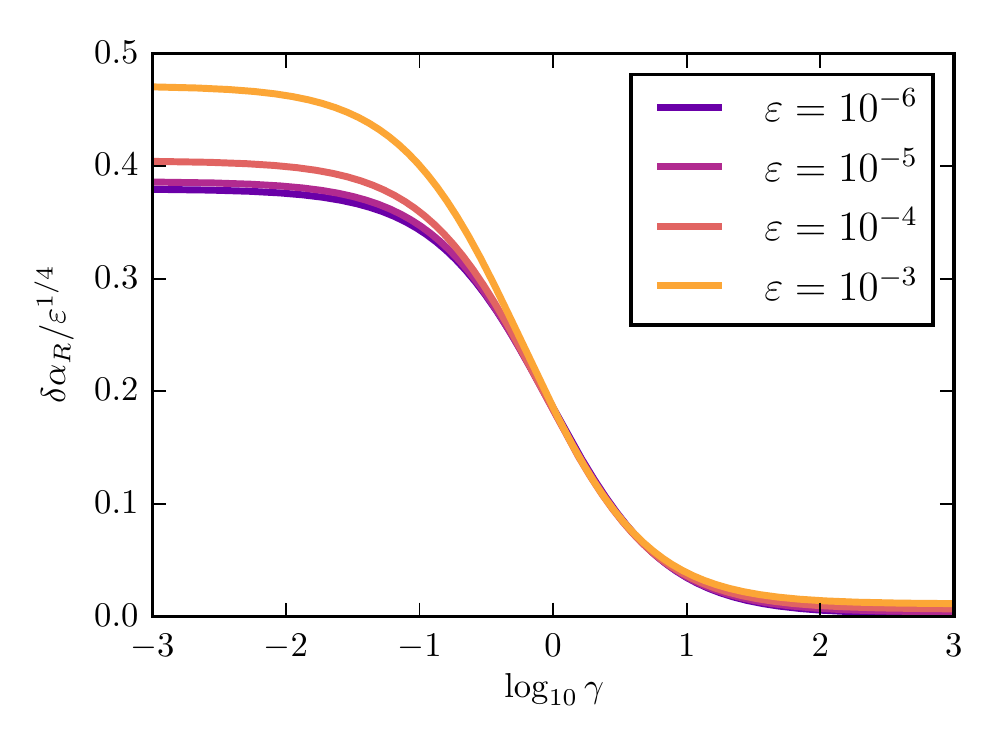}
		\caption{$\delta\alpha_R$ scaled by $\eps^{1/4}$ versus $\gamma$ for various values of $\epsilon$.}
		\label{delta_ar_sc}
	\end{center}
\end{figure}

The amplitude of the error scales with ${1-\alpha_R\propto\eps^{1/4}}$ as we can see in the Figure~\ref{delta_ar_sc}.
We plot in this Figure~\ref{delta_ar_sc} the quantity $\delta\alpha_R/\eps^{1/4}$; we see that the curves are almost similar, particularly for the smaller values of $\eps$.

\begin{table*}
	 \renewcommand\thetable{E.1}
	\begin{center}
		\caption[AMD-stability coefficients computed for the systems affected by the MMR overlap criterion]{AMD-stability coefficients computed for the systems affected by the MMR overlap criterion}

		\label{table_betas}
		%\begin{table}[ht]
\begin{tabular}{rl*{5}{D{.}{.}{3}}}
\hline\\ 
 Planet & \multicolumn{1}{l}{Period (d)} & \multicolumn{1}{l}{Mass ($\mathcal{M}^\mathrm{N}_\mathrm{E}$)} & \multicolumn{1}{l}{Eccentricity} & \multicolumn{1}{l}{$\sqrt{\langle e^2\rangle}$} &\multicolumn{1}{l}{$\beta$} &\multicolumn{1}{l}{$\beta^{(\mathrm{MMR})}$}\\ 
\hline
HD 128311&\multicolumn{6 }{l}{Mass: 0.84 $M^\mathrm{N}_\odot$}\\ 
\hline
b&454.2&463.14&0.345&0.352&0.312& \\ 
c&923.8&1032.46&0.230&0.244&3.200&27.931\\ 
\hline
HD 200964&\multicolumn{6 }{l}{Mass: 1.44 $M^\mathrm{N}_\odot$ }\\ 
\hline
b&613.8&587.98&0.040&0.067&0.024& \\ 
c&825&284.46&0.181&0.184&3.872& \multicolumn{1}{r}{$+\infty$} \\ 
\hline
HD 204313&\multicolumn{6 }{l}{Mass: 1.045 $M^\mathrm{N}_\odot$}\\ 
\hline
c&34.905&17.58&0.155&0.184&16.664& \\ 
b&2024.1&1360.31&0.095&0.095&0.110&0.110\\ 
d&2831.6&533.95&0.280&0.308&8.032& \multicolumn{1}{r}{$+\infty$} \\ 
\hline
HD 33844&\multicolumn{6 }{l}{Mass: 1.75 $M^\mathrm{N}_\odot$}\\ 
\hline
b&551.4&622.94&0.150&0.180&0.084& \\ 
c&916&556.20&0.130&0.189&2.939&22.676\\ 
\hline
HD 45364&\multicolumn{6 }{l}{Mass: 0.82 $M^\mathrm{N}_\odot$}\\ 
\hline
b&226.93&59.50&0.168&0.171&0.070& \\ 
c&342.85&209.10&0.097&0.099&1.975&13.700\\ 
\hline
HD 47366&\multicolumn{6 }{l}{Mass: 1.81 $M^\mathrm{N}_\odot$}\\ 
\hline
b&363.3&556.20&0.089&0.138&0.146& \\ 
c&684.7&591.16&0.278&0.292&2.896&2.896\\ 
\hline
HD 5319&\multicolumn{6 }{l}{Mass: 1.56 $M^\mathrm{N}_\odot$}\\ 
\hline
b&675&616.59&0.120&0.162&0.053& \\ 
c&886&365.50&0.150&0.171&8.659& \multicolumn{1}{r}{$+\infty$} \\ 
\hline
HD 73526&\multicolumn{6 }{l}{Mass: 1.08 $M^\mathrm{N}_\odot$}\\ 
\hline
b&188.9&715.11&0.290&0.293&0.200& \\ 
c&379.1&715.11&0.280&0.289&3.391&9.922\\ 
\hline
\end{tabular}%\end{table}
\\
		\textbf{Note:} Masses are given in terms of nominal terrestrial masses $ \mathcal{M}^\mathrm{N}_\mathrm{E}$ and stellar masses in terms of nominal solar masses $ \mathcal{M}^\mathrm{N}_\odot$ as recommended by the IAU 2015 Resolution B3 \citep{Prsa2016}.
	\end{center}
\end{table*}
\section{AMD-stability coefficients of the system affected by the MMR overlap criterion}

We report in Table \ref{table_betas} the AMD-stability coefficients of the systems where more than 5\% of the Monte Carlo realizations were affected by the change of critical AMD.
Apart for the system HD~47366 where 16\% of the simulations used the new criterion, the seven other systems used the critical AMD $\Cce$ for almost all the realizations.
For HD~204313, only the pair (b/d) is affected.

In Table \ref{table_betas}, $\sqrt{\langle e^2\rangle}$ corresponds to the mean value of the squared eccentricity computed as explained in section (\ref{methodo}).

\end{document}